\documentclass{aastex63}

\usepackage{multirow}
\received{}
\revised{}
\accepted{}
\submitjournal{APJ}

\shorttitle{Type IV bursts in solar cycle 24}
\shortauthors{Kumari et al.}

\begin{document}

\title{On the occurrence of type IV solar radio bursts in the solar cycle 24 and their association with coronal mass ejections}

\correspondingauthor{Anshu Kumari}
\email{anshu.kumari@helsinki.fi}

\author[0000-0001-5742-9033]{Anshu Kumari}
\affiliation{Department of Physics, \\
University of Helsinki, P.O. Box 64, \\
FI-00014 Helsinki, Finland}

\author[0000-0002-8416-1375]{D. E. Morosan}
\affiliation{Department of Physics, \\
University of Helsinki, P.O. Box 64, \\
FI-00014 Helsinki, Finland}

\author[0000-0002-4489-8073]{E. K. J. Kilpua}
\affiliation{Department of Physics, \\
University of Helsinki, P.O. Box 64, \\
FI-00014 Helsinki, Finland}



\begin{abstract}

Solar activity, in particular coronal mass ejections (CMEs), are often accompanied by bursts of radiation at metre wavelengths. Some of these bursts have a long duration and extend over a wide frequency band, namely, type IV radio bursts. However, the association of type IV bursts with coronal mass ejections is still not well understood. In this article, we perform the first statistical study of type IV solar radio bursts in the solar cycle 24. Our study includes a total of 446 type IV radio bursts that occurred during this cycle. Our results show that a clear majority, $\sim 81 \%$ of type IV bursts, were accompanied by CMEs, based on a temporal association with white-light CME observations.
However, we found that only $\sim 2.2 \%$ of the CMEs are accompanied by type IV radio bursts. We categorised the type IV bursts as moving or stationary based on their spectral characteristics and found that only $\sim 18 \%$ of the total type IV bursts in this study were moving type IV bursts. 
Our study suggests that type IV bursts can occur with both `Fast' ($\geq 500$ km/s) and `Slow' ($< 500$ km/s), and also both `Wide' ($\geq 60^{\circ}$) and `Narrow' ($< 60^{\circ}$) CMEs. However, the moving type IV bursts in our study were mostly associated with `Fast' and `Wide' CMEs ($\sim 52 \%$), similar to type II radio bursts. Contrary to type II bursts, stationary type IV bursts have a more uniform association with all CME types.


\end{abstract}



\section{Introduction}
     \label{sec:section1} 

{Coronal mass ejections are large eruptions of magnetised plasma from the Sun \citep{webb2012coronal} that are often accompanied by radio emission \citep{white2007solar}, generated by the energetic electrons produced during these eruptions \citep{gopalswamy2004intensity}. These electrons can generate radio emission in the corona through various emission mechanisms \citep{melrose1980emission}. The most common radio bursts associated with CMEs are type II and type IV bursts (see for details, \citealt{Gergely1986, Clasen2002, Vasanth_2016, Anshu2017a, kumari2017b, kumari2019direct, Morosan2020}; and the references therein). }

{CMEs are often accompanied by broadband continuum emission at decimetric and metric wavelengths, known as type IV radio bursts \citep{pick1986observations}, that can have either stationary or moving sources and various emission mechanisms \citep[for a review, see][]{Bastian1998}. Moving type IV radio bursts (hereinafter referred to as type IVm)} were first classified by \citet{Boischot1957} as broadband radiation propagating away from the Sun. This emission was believed to originate due to synchrotron or gyro-synchrotron emitting electrons, gyrating inside helical magnetic fields within the CME flux rope \citep{Boischot1968, Dulk1973}. Stationary type IV radio bursts (hereinafter referred to as type IVs) have also since been discovered originating due to plasma emission \citep{Weiss1963, Benz1976, salas2020polarisation} and, at the same time, the plasma emission mechanism has also been attributed to some type IVm bursts \citep{Gary1985, Vasanth2019, Morosan2020}. A type IV burst forming a `radio CME' was first reported by \citet{Bastian2001}, as an ensemble of CME radio loops imaged by the Nan{\c c}ay Radioheliograph. A similar `radio CME' has later been reported by \citet{Maia2007}. These loops are believed to be made visible at radio wavelengths due to synchrotron emitting electrons.

{Of particular interest in the studies of type IV radio bursts is their potential to provide estimates of the CME magnetic field strength \citep[for example,][]{Gopal1987, Bastian2001, Ramesh2004, Maia2007, Bain2014, Hariharan2016a, Carley2017}. The magnetic field dictates the dynamics of the CME eruption and its space weather consequences \citep[for example,][]{Kilpua2017}, but it has rarely been measured remotely and most measurements of the CME magnetic field rely on in situ observations at 1~AU \citep[for example,,][]{Kilpua2019}. Remote type IV observations have been used to estimate the magnetic field strength in CMEs, some of which found a range of values, for example, $<$1~G at 2~R$_\odot$ \citep{Maia2007}, 5--15~G at 1.5~R$_\odot$ \citep{Tun2013}, and 1~G at 2.2~R$_\odot$ \citep{Hariharan2016a}. However, there are currently some difficulties in explaining type IV emission as this radiation can be emitted by various emission mechanisms and sometimes multiple emission mechanisms contribute to the observed continuum emission \citep{Morosan2019}. The association of CMEs with type IV bursts is also poorly understood. Some statistical studies were carried out in previous solar cycles \citep{Gergely1986, Robinson1978}. \cite{Robinson1978} studied 23 type IV bursts with Culgoora spectropgraph and proposed a new model involving synchrotron radiation from electrons to overcome the limitations of the plasmoid theory \citep{Dulk1973, schmahl1973}. \cite{Gergely1986} found that approximately one-third to one-half of the type IV bursts observed were associated with CMEs, however this study was carried out before the availability of modern spacecraft observations. Therefore, more detailed statistical studies on the relationship between type IV emission and their association with CMEs are required. Extensive databases of both radio bursts and CMEs from ground and space based observatories over an entire solar cycle can now facilitate this type of study. Statistical studies have already been carried out on other types of radio bursts associated with CMEs, for example, type II bursts \citep{lara2003statistical, Kahler2019}, showing that these bursts are almost always associated with `Fast and Wide' CMEs. However, there is no such association made yet in the case of type IV bursts.} 

{In this article we perform a statistical study on the association of type IV radio bursts to CMEs in solar cycle 24. The paper is organised as follows: 
Section~\ref{sec:section2} presents the observational data used for this work.
The detailed data analysis methods are presented in Section~\ref{sec:section3}. Section \ref{sec:section4} contains the results of this analysis which are further discussed in Section~\ref{sec:section5}.}

\section{Observations}
\label{sec:section2}

\subsection{CME Observations}
\label{sec:section2.1}

In this study, we use CMEs detected using the Large Angle and Spectrometric Coronagraph (LASCO) onboard the Solar and Heliospheric Observatory (SOHO; \citet{Brueckner1995}) and the Cor1 and Cor2 coronagraphs from the Sun-Earth Connection Coronal and Heliospheric Investigation (SECCHI; \citet{Howard2008}) onboard the Solar Terrestrial Relationship Observatory (STEREO).
The SOHO spacecraft orbits at the Earth's L1 Lagrangian point. LASCO is a set of three coronagraphs, which observes the Sun from 1.1-3~R$_\odot$ (LASCO-c1), 2.5-6~R$_\odot$ (LASCO-c2) and 4-32~R$_\odot$ (LASCO-c3). LASCO-c2 and LASCO-c3 are currently operational with a spatial resolution of $11.4^{'}$ and of $56^{'}$, respectively. The image time cadence is $\sim$ 12 min for both the coronagraphs.
We used the CMEs listed in the Coordinated Data Analysis Workshop (CDAW) CME database\footnote{\url{https://cdaw.gsfc.nasa.gov/}}. CDAW is an online CME catalogue \citep{Yashiro2004,Yashiro2008,Gopalswamy2009cdaw} which contains manual detections of CMEs observed with LASCO. The CME catalogue contains CMEs identified since the launch of SOHO in 1996 and it includes CME properties such as speed, position angle, acceleration, mass, and kinetic energy, along with CME plots and movies\footnote{\url{https://cdaw.gsfc.nasa.gov/CME_list/index.html}}.

The STEREO spacecrafts are two identical spacecrafts, Ahead (A) and Behind (B), orbiting the Sun ahead and behind the Earth, respectively. SECCHI has two white-light coronagraphs, Cor1 and Cor2, which observe the Sun from 1.3-4~R$_\odot$ and 2-15~R$_\odot$, respectively. These coronagraphs provide polarised maps of high temporal and spatial resolution ($7.5^{'}$ and $15^{'}$, respectively).   
For Cor1, we used the CME detection provided by the Goddard Space Flight Center(GFSC) Cor1 website\footnote{\url{https://cor1.gsfc.nasa.gov/catalog/}}.
For Cor2, we used the CME list provided on the Solar Eruptive Event Detection System (SEEDS)\footnote{\url{http://spaceweather.gmu.edu/seeds/secchi/detection_cor2/monthly/}} website, which automatically detects solar eruptions. We also used the list of CMEs detected in Cor2 coronagraphs of STEREO-A and STEREO-B provided by
\citet{Vourlidas2017}\footnote{
\url{http://solar.jhuapl.edu/Data-Products/COR-CME-Catalog.php}}.

\subsection{Radio Observations}
\label{sec:section2.2}

The occurrence of radio bursts is provided by the event lists from the Space Weather Prediction Center\footnote{\url{https://www.swpc.noaa.gov/products/solar-and-geophysical-event-reports}}, which catalogues radio events since the year 1996\footnote{\url{ftp://ftp.swpc.noaa.gov/pub/warehouse/}}. SWPC collects the reports from contributing stations, compiles them, and updates the list every $\sim$ 30 minutes. This list is manually reviewed and it includes radio bursts, optical and X-ray flares, and properties of these emissions. The contributing observatories are Culgoora spectrograph\footnote{\url{https://www.sws.bom.gov.au/Solar/2/2}}, Learmonth spectrograph\footnote{\url{https://www.sws.bom.gov.au/Solar/3/2}}, optical observatories, for example, Holloman Solar Observatory\footnote{\url{https://www.holloman.af.mil/News/Display/Article/1801805/holloman-solar-observatory/}} and GOES satellite data\footnote{\url{https://www.goes.noaa.gov/}}. 


\section{Data Analysis}
\label{sec:section3}


The SWPC list described in the previous section is used to extract the type IV radio bursts and their properties (such as start time, duration, and bandwidth) during solar cycle 24. 
We divided the type IV bursts into two subgroups: i) moving type IV bursts (IVm); and ii) stationary type IV bursts (IVs), based on their spectral characteristics (drift rates and duration). 

The drift rate (DR) was calculated for type IVm bursts using the relation $DR = \frac{F_H-F_L}{t}$, where $DR$, $F_H, F_L$ and $t$ are the drift rate, start frequency, end frequency and duration of the bursts, respectively. The duration of these bursts is estimated using the relation $t = t_{end} - t_{start}$, where $t, t_{end}$ and $t_{start}$ are the duration, start time and end time of the type IV bursts, respectively. For the stationary/moving type IV bursts classification, in addition to the drift rate, we also considered the duration of the moving type IV bursts as mentioned by \cite{Robinson1978} and \cite{Gergely1986}.
These studies reported that moving type IV bursts have duration of less than an hour. We first classified the type IV bursts based on their duration. The drift rates for most of the moving bursts classified this way were $\geq 0.03$ MHz. We then manually checked the spectra of the rest of the type IV bursts (with duration $>$ 1 hour and drift rates $\geq$ 0.03 MHz to determine if they are moving or stationary bursts. After these three steps, we prepared a list of moving type IV bursts and the remaining bursts were classified as stationary type IV bursts.

\begin{figure}    
   \centering\includegraphics[width=0.49\textwidth,clip=]{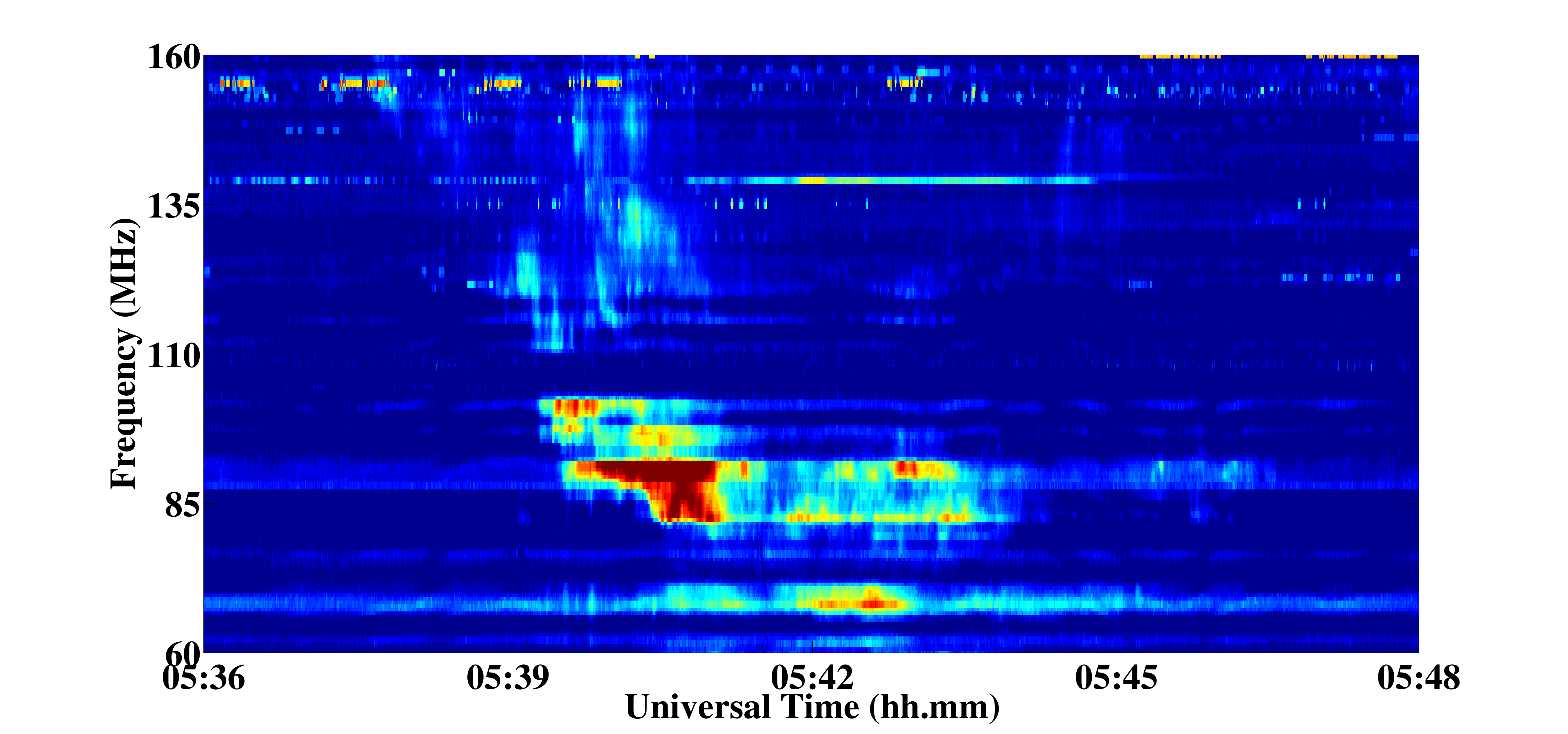}
   \centering\includegraphics[width=0.49\textwidth,clip=]{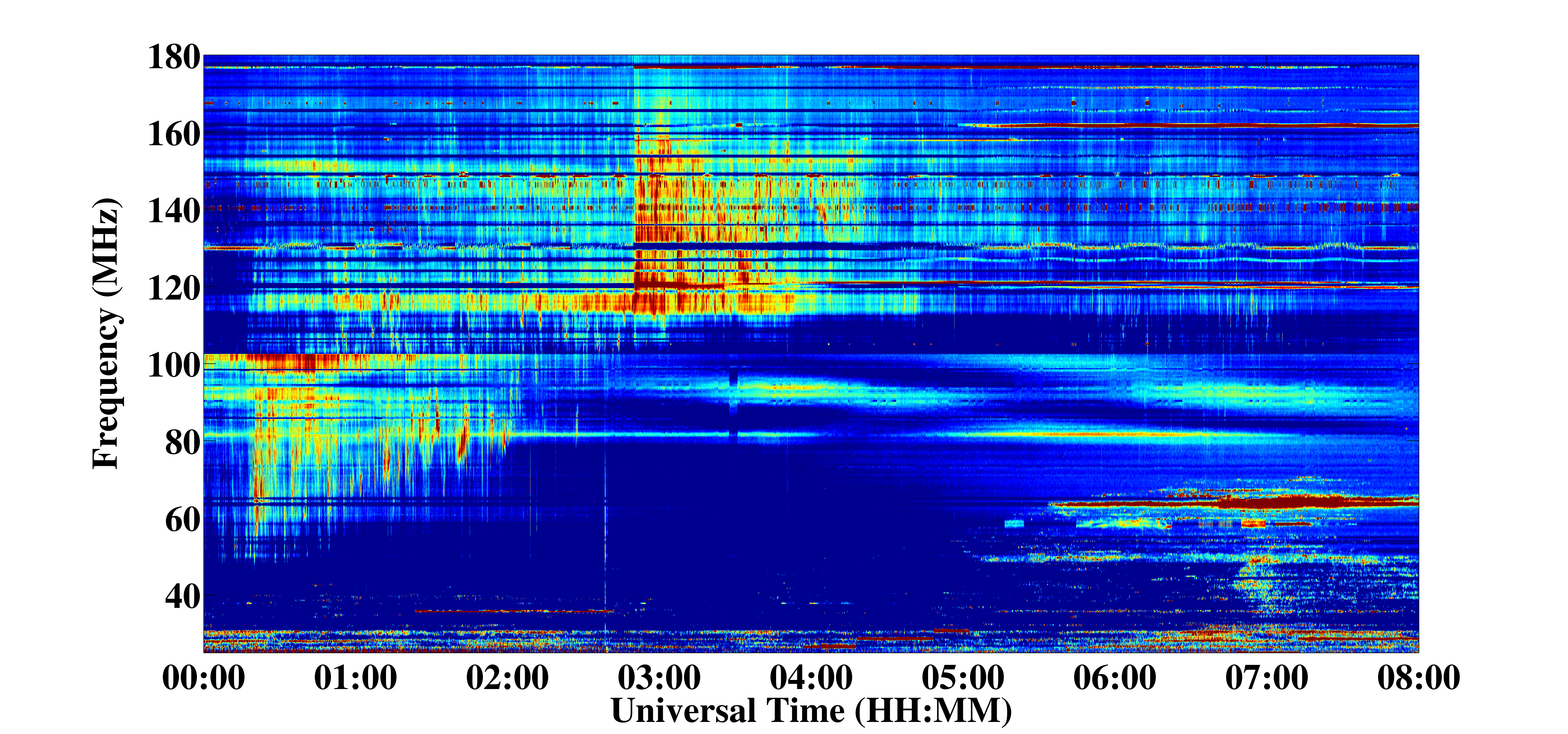}
              \caption{\textbf{Left panel:} Solar radio dynamic spectra of a moving type IV burst on October 18, 2017. The duration of the burst is $\sim 7 $ minutes and it shows a frequency drift of $\sim 0.09$ MHz/s from 160 MHz to 60 MHz.
              \textbf{Right panel:} Solar radio dynamic spectra of a stationary type IV burst on October 03, 2011. The duration of the burst is $\sim 7 $ hours. 
                      }
   \label{fig:figure1a}
   \end{figure}

Figure \ref{fig:figure1a} shows an example of a type IVm (left) and a type IVs burst (right) that occurred on 17 October, 2017\footnote{\url{http://soleil.i4ds.ch/solarradio/qkl/2017/10/18/ALMATY_20171018_053000_58.fit.gz.png}} and 03 October, 2011, respectively. Type IVm bursts generally have a large drift rate and a short duration (see for example, \citealt{melnik2018solar}). The type IVm burst in the left panel of \ref{fig:figure1a} starts at $\sim $ 5:38 UT and lasts for $\sim 7$ minutes, with a drift rate of $\sim 0.09$ MHz/s. Type IVs bursts in turn have negligible  drift rates and can last for few minutes up to a few hours (see for example, \citealt{pick1986observations}). The type IVs burst in the right panel of Figure \ref{fig:figure1a} starts at $\sim $ 00:20 UT and lasts for $\sim 7 $ hours.
We would like to note that not all the type IVm bursts are short-lived (i.e. duration $\leq$ 1 hour). For example, previous case studies by \cite{Ramesh2013} and \cite{Vasanth2019} have shown that duration of type IVm bursts were $>1.5$ and $>2.5$ hours, respectively. However, the amount reported long duration type IVm bursts are very less in this solar cycle, hence we used the duration criteria mentioned by \cite{Robinson1978} and \cite{Gergely1986} for the large number of type IV bursts (446 bursts) presented in this study. To understand the origin, emission mechanism and evolution of type IVm and IVs bursts, case studies are important (see for example, \cite{Koval2016, Liu2018, Morosan2019}; and the references therein).


In order to study the association of type IV bursts with CMEs, we extract the CMEs and their properties (such as start time, position angle, angular width and linear speed) from LASCO, STEREO-A and STEREO-B catalogues. 
Since white-light CME images use an occulting disc, the first appearance of CMEs in the LASCO C2 field of view (FOV) is expected to be later in time than the onset of the type IV bursts. The first appearance of the CME in LASCO-c2 FOV depends upon the speed and acceleration of the CME. For the CME-type IV association, we used the following criteria:
a CME should appear in coronagraph FOV within  $\approx 2 $ hours of the start time of the type IV burst (we note that time between CME and type IV bursts association can vary depending on the location of the ejection, for example, a limb CME may appear early in coronagraph images). 
Both of these criteria were chosen to assure that the correct pair of white-light CME and radio signature were connected. In the case that these criteria were not enough to determine a CME association within SOHO/LASCO-c2 and STEREO-cor2, the STEREO-cor1 coronagraph data was investigated manually.

\begin{table}[]
\caption{Number of events for 2009-2019}
\label{tab:table1}
\begin{tabular}{ccccccccc} 
\hline
Year & CMEs & $\%$ CMEs & Type IV & $\%$ Type IV  & Type IVm & $\%$ Type IVm & Type IVs & $\%$ Type IVs\\ 
\hline
2009         & 746  & 4.6 $\%$  & 1   & 0.2 $\%$  & 0   & 0.0 $\%$ & 1   & 0.3 $\%$   \\
2010    & 1117  & 6.9 $\%$ & 3   & 0.7 $\%$  & 1   & 1.2 $\%$ & 2   & 0.5 $\%$   \\
2011     & 1990   & 12.4 $\%$ & 65 & 14.6 $\%$  & 14  & 17.5 $\%$ & 51 & 13.9 $\%$    \\
2012         & 2177   & 13.5 $\%$ & 80 & 18.0 $\%$ & 14  & 17.5 $\%$ & 66   & 18.0 $\%$   \\
2013    & 2338  & 14.5 $\%$  & 83  & 18.6 $\%$ & 17  & 21.3 $\%$ & 66   & 18.0 $\%$   \\
2014    & 2478   & 15.4 $\%$   & 127 & 28.4 $\%$ & 14  & 17.5 $\%$ & 113   & 30.9 $\%$  \\
2015         & 2058    & 12.8 $\%$  & 51  & 11.5 $\%$ & 12   & 15.0 $\%$ & 39 & 10.7 $\%$   \\
2016      & 1393    & 8.7 $\%$  & 16  & 3.6 $\%$  & 3   & 3.8 $\%$ & 13   & 3.6 $\%$   \\
2017    & 786     & 4.9 $\%$  & 20  & 4.4 $\%$  & 5   & 6.2 $\%$ & 15  & 4.1 $\%$   \\
2018    & 476   &    2.9 $\%$  & 0  & 0 $\%$  & 0   & 0.0 $\%$ & 0   & 0.0 $\%$  \\
2019    & 543 & 3.4 $\%$  & 0 & 0 $\%$   & 0   & 0.0 $\%$ & 0   & 0.0 $\%$  \\
All    & 16107 & 100.0 $\%$  & 446 & 100.0 $\%$   &  80  & 17.9 $\%$ & 366   & 82.1 $\%$  \\
\hline
\end{tabular}
\end{table}

 \begin{figure}    
   \centering\includegraphics[width=0.32\textwidth,clip=]{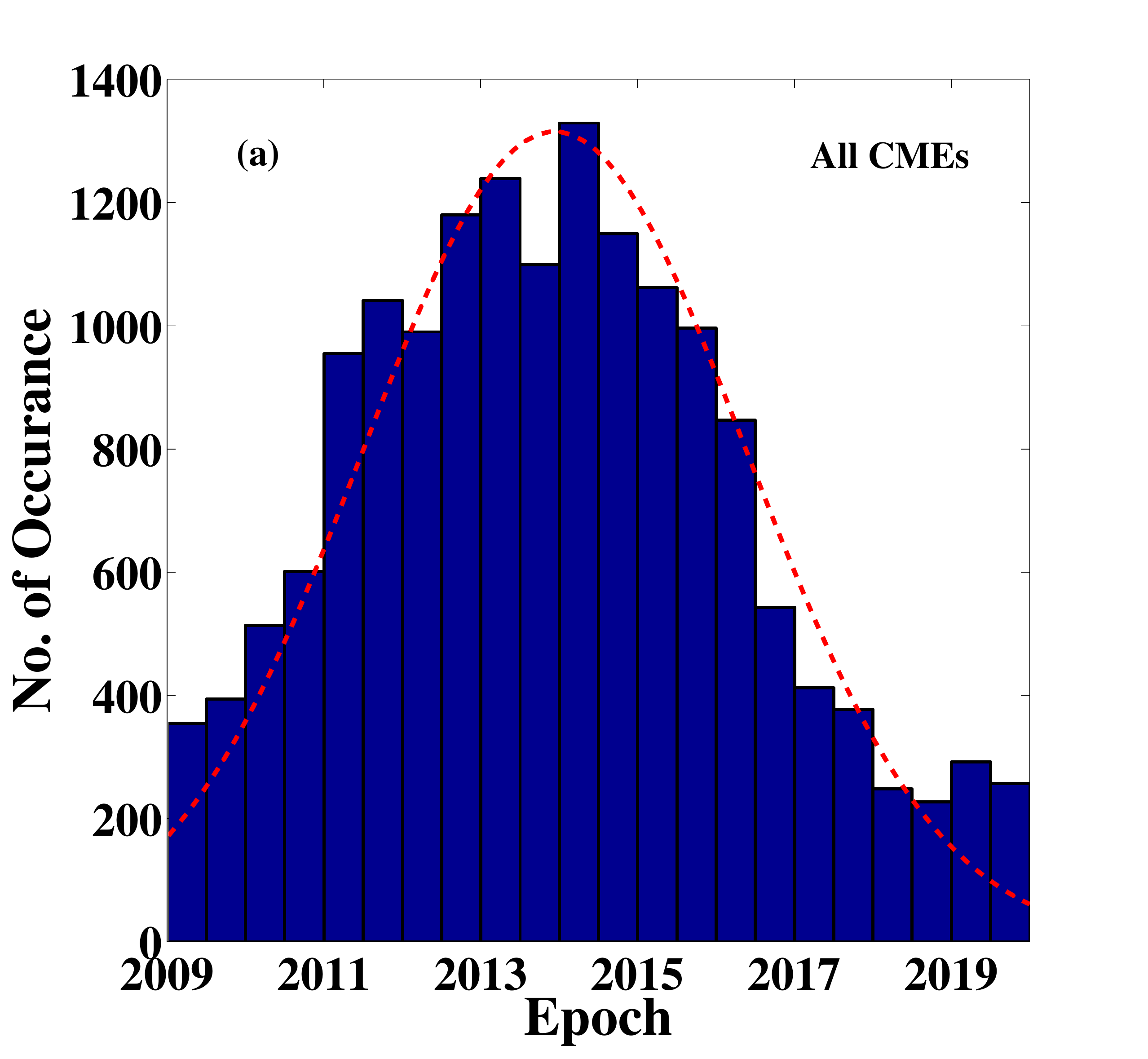}
   \centering\includegraphics[width=0.32\textwidth,clip=]{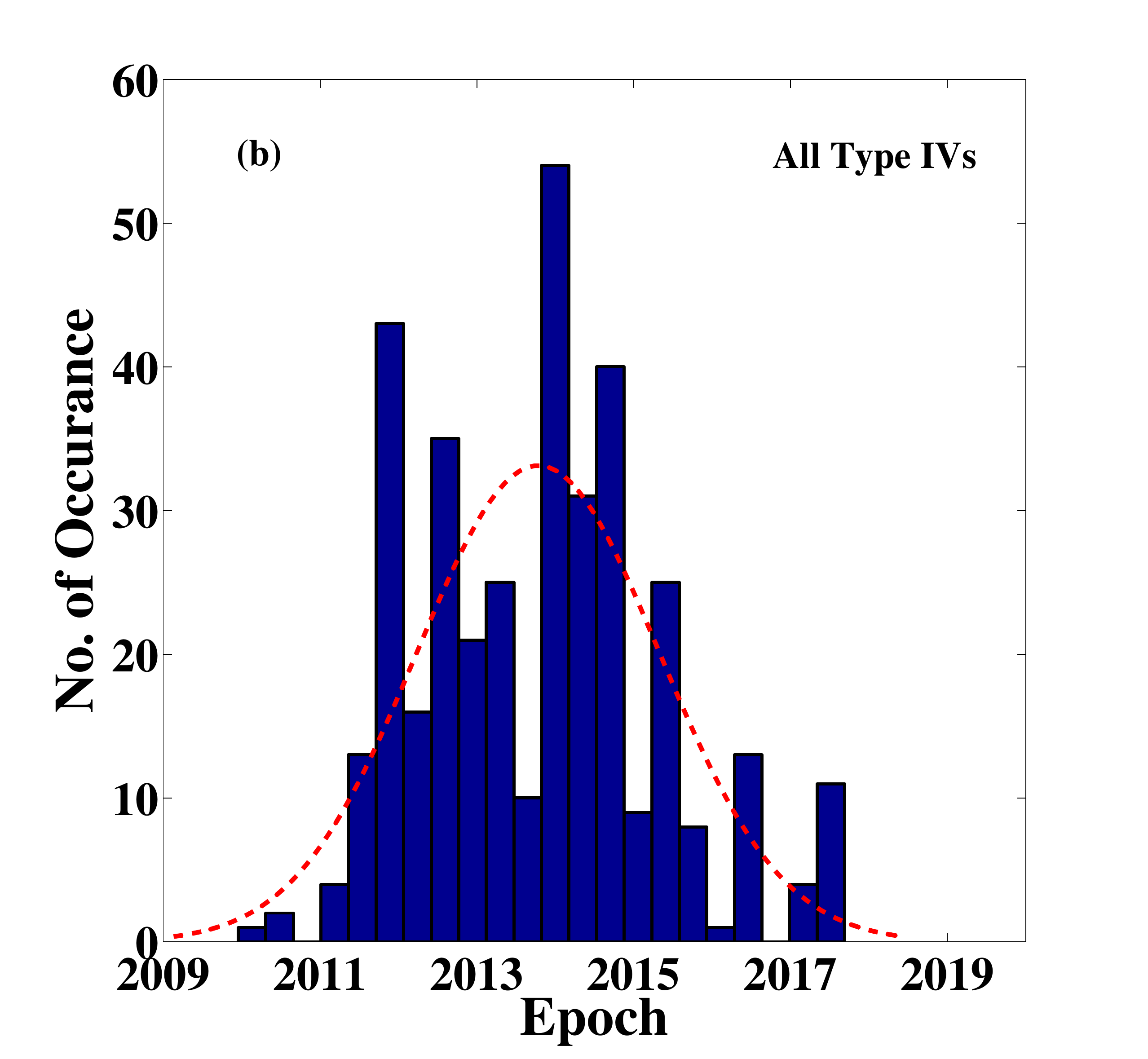}
      \centering\includegraphics[width=0.32\textwidth,clip=]{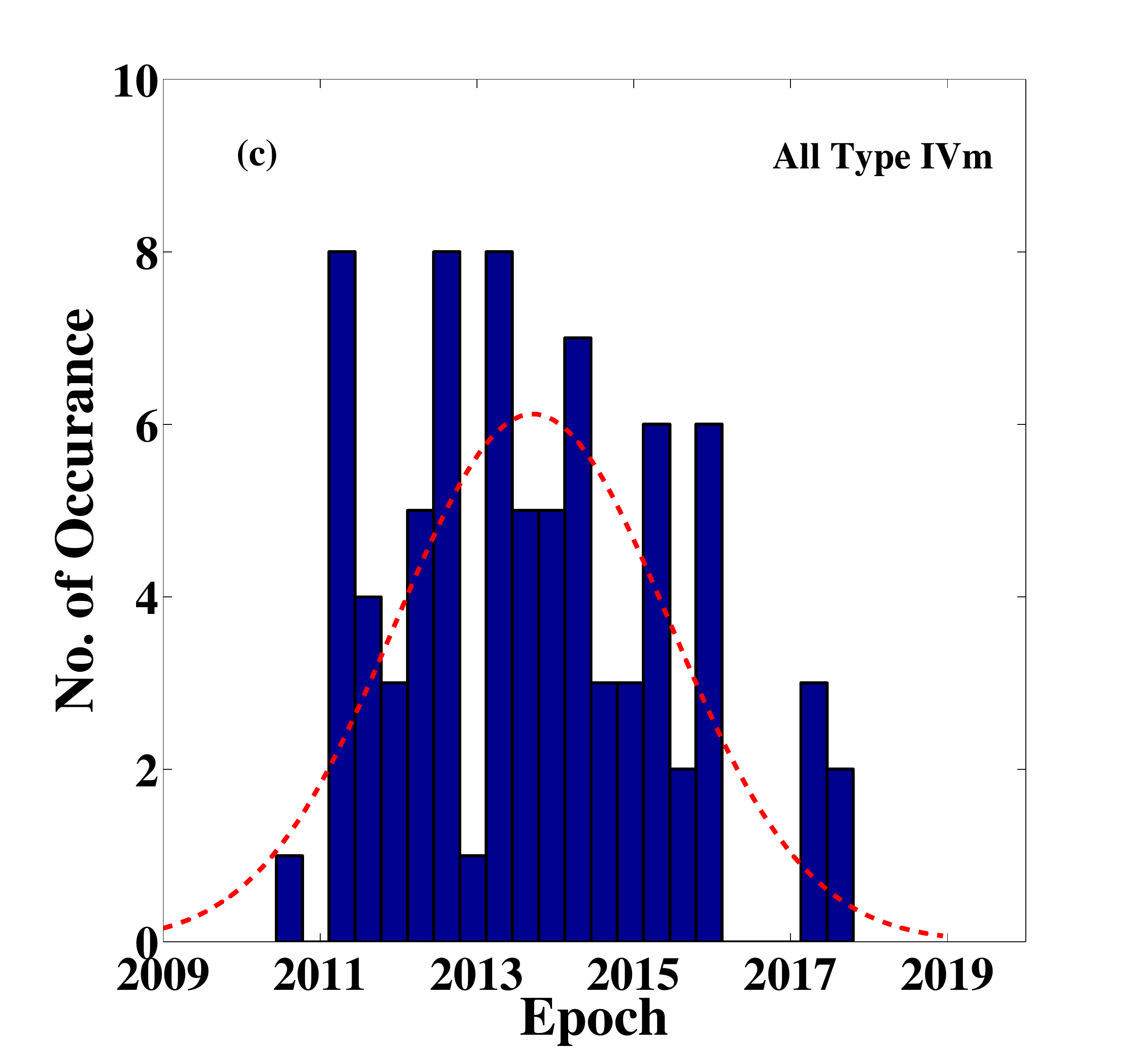}
   \centering\includegraphics[width=0.32\textwidth,clip=]{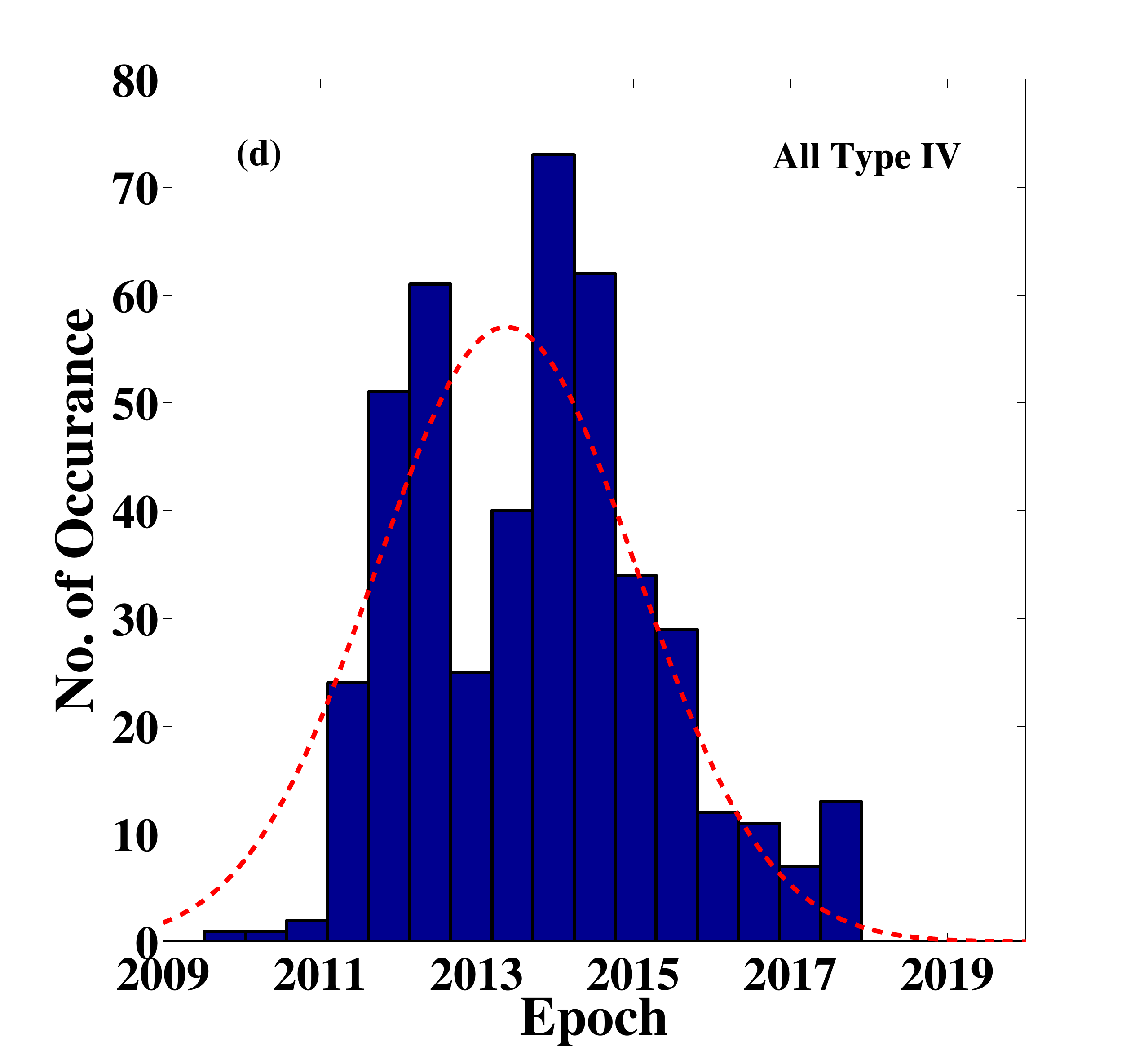}
      \centering\includegraphics[width=0.32\textwidth,clip=]{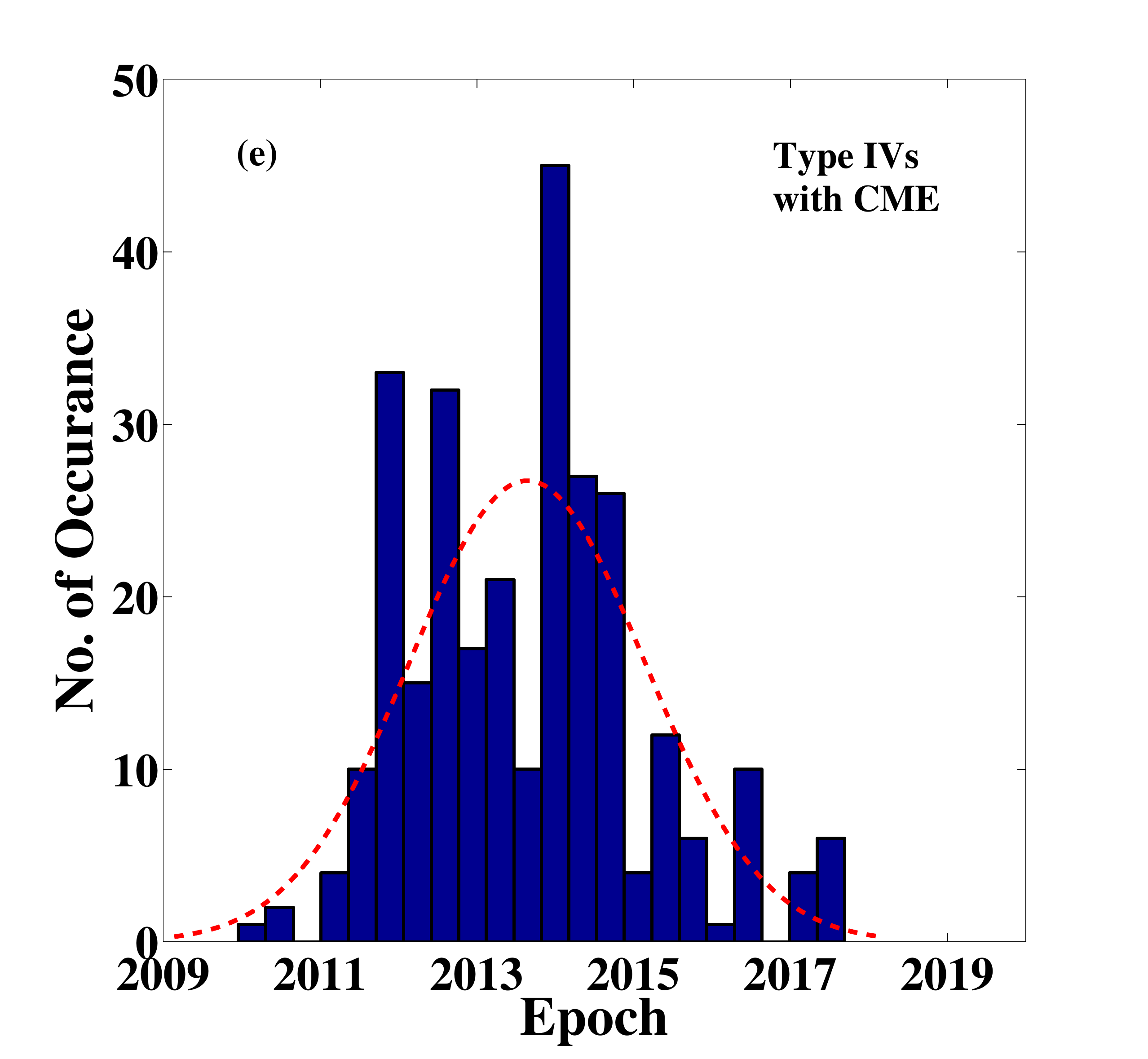}
   \centering\includegraphics[width=0.32\textwidth,clip=]{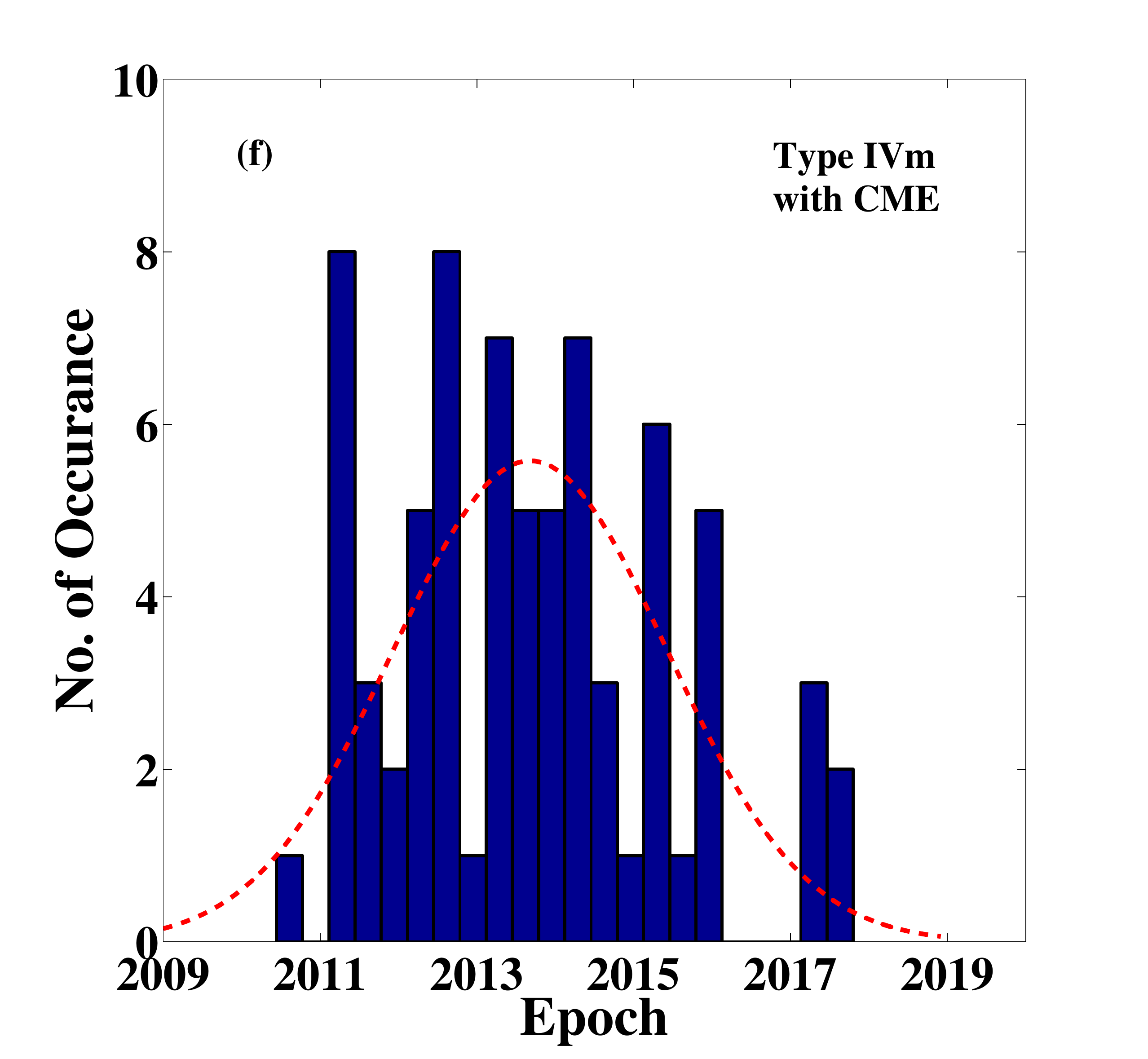}
              \caption{ The distribution of CMEs and type IV bursts in the solar cycle 24. Histogram for 
              (a) all CMEs;
              (b) all type IVs bursts;
              (c) all type IVm bursts;
              (d) all type IV bursts;
              (e) all type IVs bursts with CMEs;
              (f) all type IVm bursts with CMEs;
                      }
   \label{fig:figure1}
   \end{figure}

\section{Results}
\label{sec:section4}

{The total number of CMEs and type IV bursts observed in Solar Cycle 24 is shown in Table \ref{tab:table1}. Figure \ref{fig:figure1} shows the histogram of occurrence of CMEs (from LASCO only) and type IV bursts in solar cycle 24. 
There were total 16107 CMEs reported with SOHO/LASCO and additional 5944 with STEREO-A and B combined.
Figure \ref{fig:figure1} indicates that the CME occurrence rate follows a Gaussian distribution with the peak during the maximum of this solar cycle (Jan--June, 2014). Our analysis shows that there were $\sim 60\%$ more CMEs during the rising phase compared to the declining phase of the cycle. The majority of CMEs, $\sim 66\%$ occurred during the peak years of this cycle (2012-2016).}

In total, 446 type IV bursts were observed in this cycle, of which $\sim 82\%$ (366) were stationary (type IVs) and the remaining $\sim 18\%$ (80) were moving (type IVm). From all type IV bursts included in this study, 359 type IV bursts ($\sim 81\%$) showed a temporal and spatial correlation with CMEs, while only $\sim 2.2\%$ of the CMEs were accompanied by type IV bursts. 
There were no type IV bursts observed during the start and end years of the solar cycle 24, i.e 2009, 2018 and 2019. These were the years of the lowest solar activity. Figure \ref{fig:figure1} and Table \ref{tab:table1} indicate that the distribution of occurrence of type IV bursts, both type IVm and IVs, also follow approximately a Gaussian distribution with peaks during January to June 2014, but with larger year to year variations.


{We categorised all CMEs based on their linear speeds and angular widths as `Fast', `Slow', `Wide' and `Narrow' CMEs. CMEs with linear speed $\geq 500$ km/s are classified as `Fast' CMEs, with the remaining classified as `Slow'. Similarly the CMEs with angular width $\geq 60^{\circ}$ are classified as `Wide' CMEs, with the remaining classified as `Narrow' CMEs. Table \ref{tab:table2} contains the list of CMEs and their properties based on their linear speed and width. Most of the CMEs in this solar cycle were `Slow and Narrow' ($\sim 65 \%$). }

     \begin{table}[]
\caption{Occurrence and association of CMEs with type IV bursts based on their speed and width: Fast (linear speed $\geq$ 500 km/s), Slow (linear speed $<$ 500  km/s) Wide (angular width $\geq 60^{\circ}$ ) and Narrow (angular width $< 60^{\circ})$ CMEs}
\label{tab:table2}
\begin{tabular}{ccccccc}
\hline
Category         & CMEs & $\% $ CMEs & Type IVm & $\% $ Type IVm  & Type IVs & $\% $ Type IVs \\
                 \hline  
Fast             & 2473        & 15.4 $\%$  & 45 & 61.6   $\%$   & 132 & 46.2   $\%$  \\
Slow               & 13634       & 85.6   $\%$  & 28      & 38.4   $\%$  & 154 & 53.8   $\%$  \\
Wide               & 4345        & 27.0    $\%$   & 51  & 69.9    $\%$  & 168 & 58.7   $\%$ \\
Narrow            & 11762     & 73.0   $\%$  & 22       & 30.1   $\%$ & 118 & 41.3   $\%$    \\
Fast $\&$ Wide     & 1216        & 7.6    $\%$  & 38 & 52.1    $\%$  & 102 & 35.7   $\%$   \\
Fast $\&$ Narrow   & 1257        & 7.8    $\%$  & 7 &  9.6  $\%$  & 30 & 10.5   $\%$   \\
Slow $\&$ Wide      & 3129        & 19.4   $\%$   & 13  & 17.8$\%$  & 66 & 23.1   $\%$ \\
Slow $\&$ Narrow   & 10505       & 65.2     $\%$ & 15   & 20.5   $\%$ & 88 & 30.7   $\%$ \\
All  & 16107     & 100.0   $\%$ & 73   & 100.0   $\%$ & 286 & 100.0   $\%$ \\
\hline
\end{tabular}
\end{table}

\subsection{CMEs associated with moving type IV bursts and their properties}
\label{sec:section4.2}

\begin{figure}    
   \centerline{\includegraphics[width=.9\textwidth, clip=]{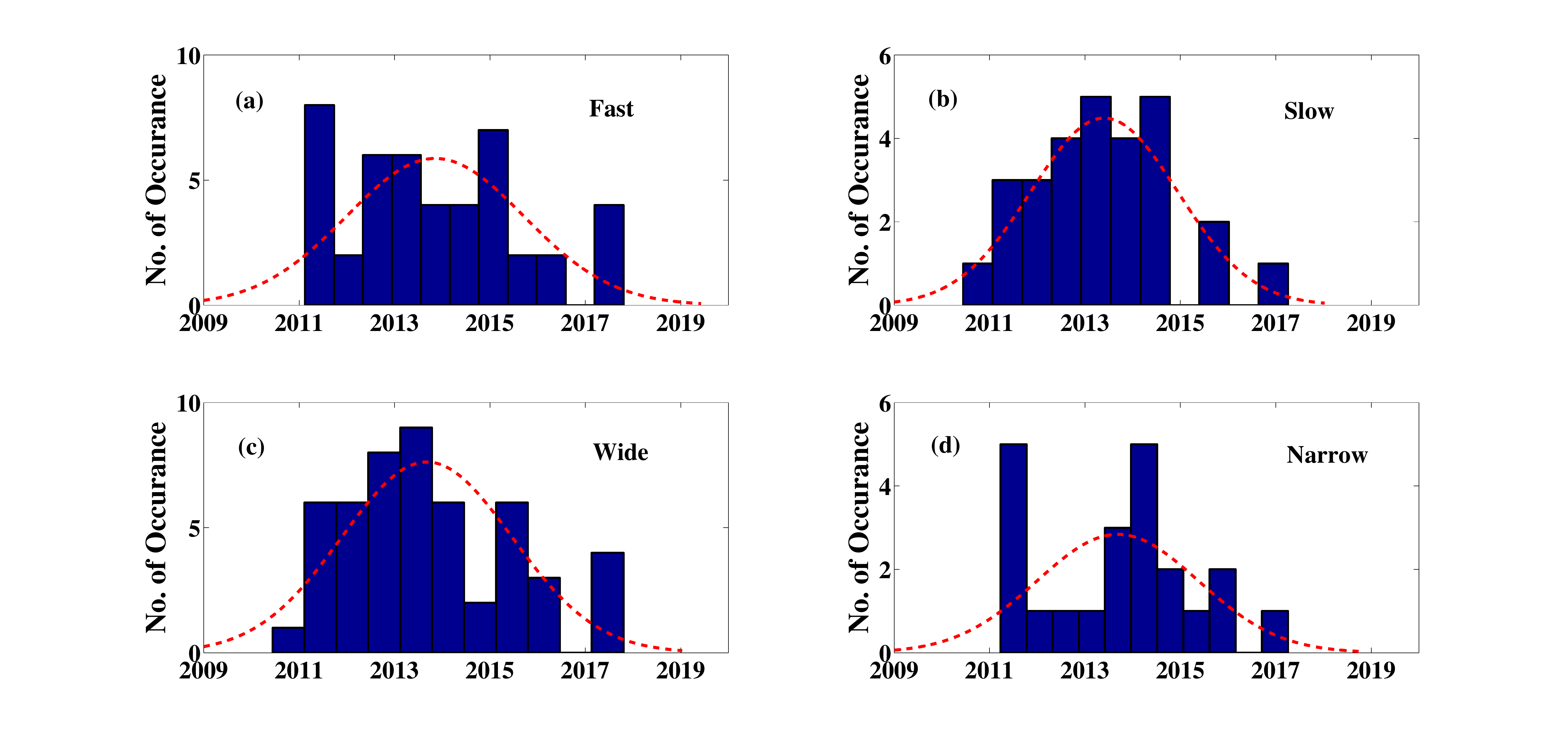}}
   \centerline{\includegraphics[width=.9\textwidth,clip=]{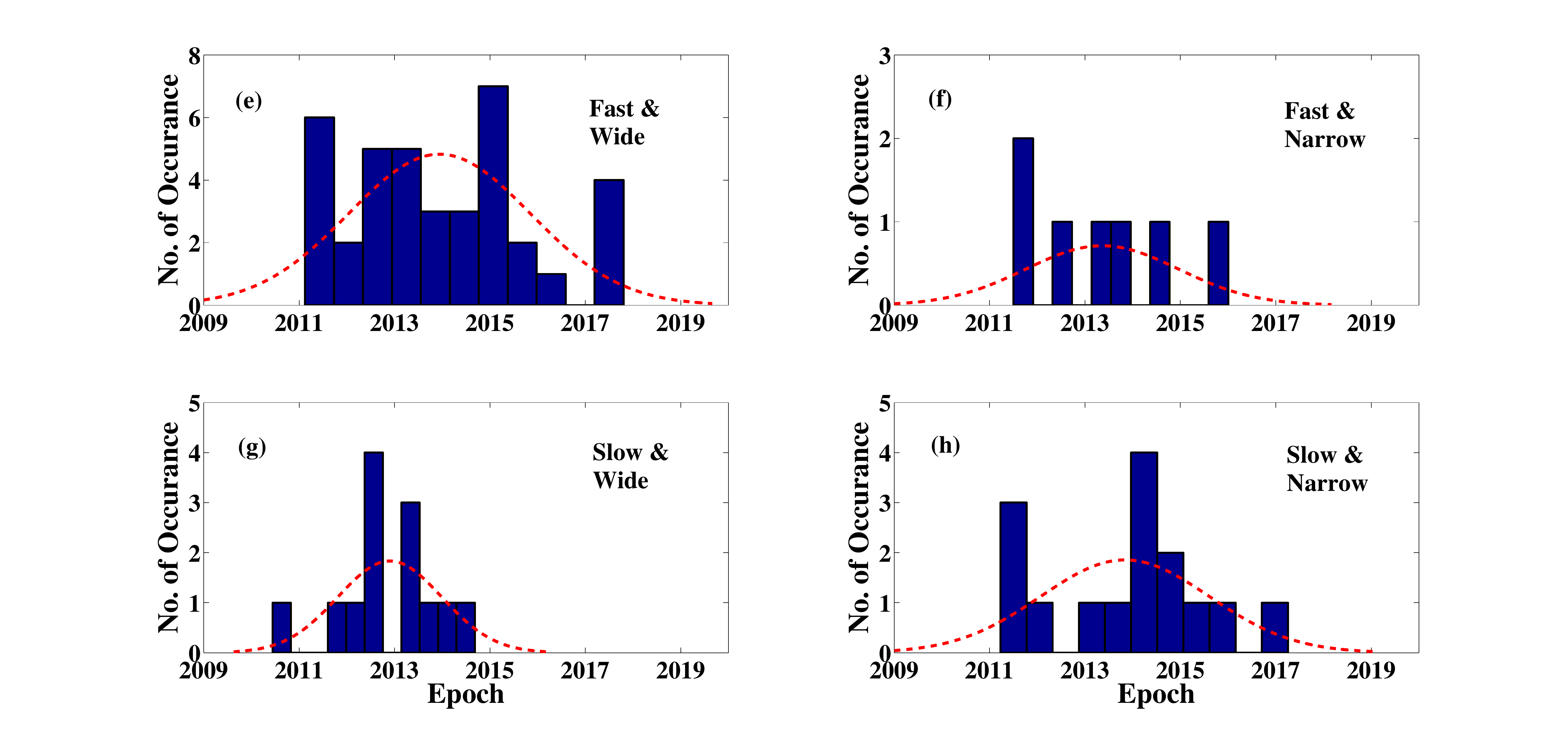}}
              \caption{  The variation of angular width and linear speed of CMEs which were accompanied with type IVm bursts in the solar cycle 24. Histogram for
              (a) Fast CMEs;
              (b) Slow CMEs;
              (c) Wide CMEs;
              (d) Narrow CMEs;
              (e) Fast and Wide CMEs;
              (f) Fast and Narrow CMEs;
              (g) Slow and Wide CMEs;
              (h) Slow and Narrow CMEs.
                      }
   \label{fig:figure5a}
   \end{figure}

A distribution of the moving type IV bursts (type IVm) associated with CMEs in solar cycle 24 is shown in Figure~\ref{fig:figure1}(f). 
Out of 80 type IVm bursts observed, 73 ($\sim 91 \%$) were accompanied by white-light CMEs. However, only $\sim 0.5\%$ of the total CMEs in this solar cycle were accompanied by type IVm bursts. 
Table \ref{tab:table2} contains the list of type IVm bursts which were associated with `Fast', `Slow', `Wide' and `Narrow' CMEs. Of these, $\sim62\%$ were associated with `Fast' CMEs and $\sim38\%$ were associated with `Slow' CMEs, unlike type II radio bursts where the majority are associated with `Fast' CMEs \citep{Kahler2019}. Similar to type II bursts \citep{Kahler2019}, `Wide' CMEs are more likely to be related to type IVm bursts than 'Narrow' CMEs, with associations $\sim70\%$ and $\sim30\%$, respectively. Figure \ref{fig:figure5a} shows histograms of the distribution of type IVm bursts associated with various CME types. The spread of type IV bursts in this case also follows a Gaussian distribution with peaks during January to June 2014 (during the solar maximum) for `Fast', `Slow', `Wide' and `Narrow' CMEs. For a combination of speed and width of CMEs, there are no clear trends (see Figure \ref{fig:figure5a} bottom panel histograms). 
Our analysis shows that over half of the type IVm bursts ($\sim52\%$) in this solar cycle were associated with `Fast' and `Wide' CMEs. 
In turn, there were very few moving type IV bursts ($\sim 10 \%$) associated with `Fast and Narrow' CMEs (see Figure \ref{fig:figure5a}f).

\subsection{CMEs associated with stationary type IV bursts and their properties}
\label{sec:section4.3}

\begin{figure}    
   \centerline{\includegraphics[width=.9\textwidth, clip=]{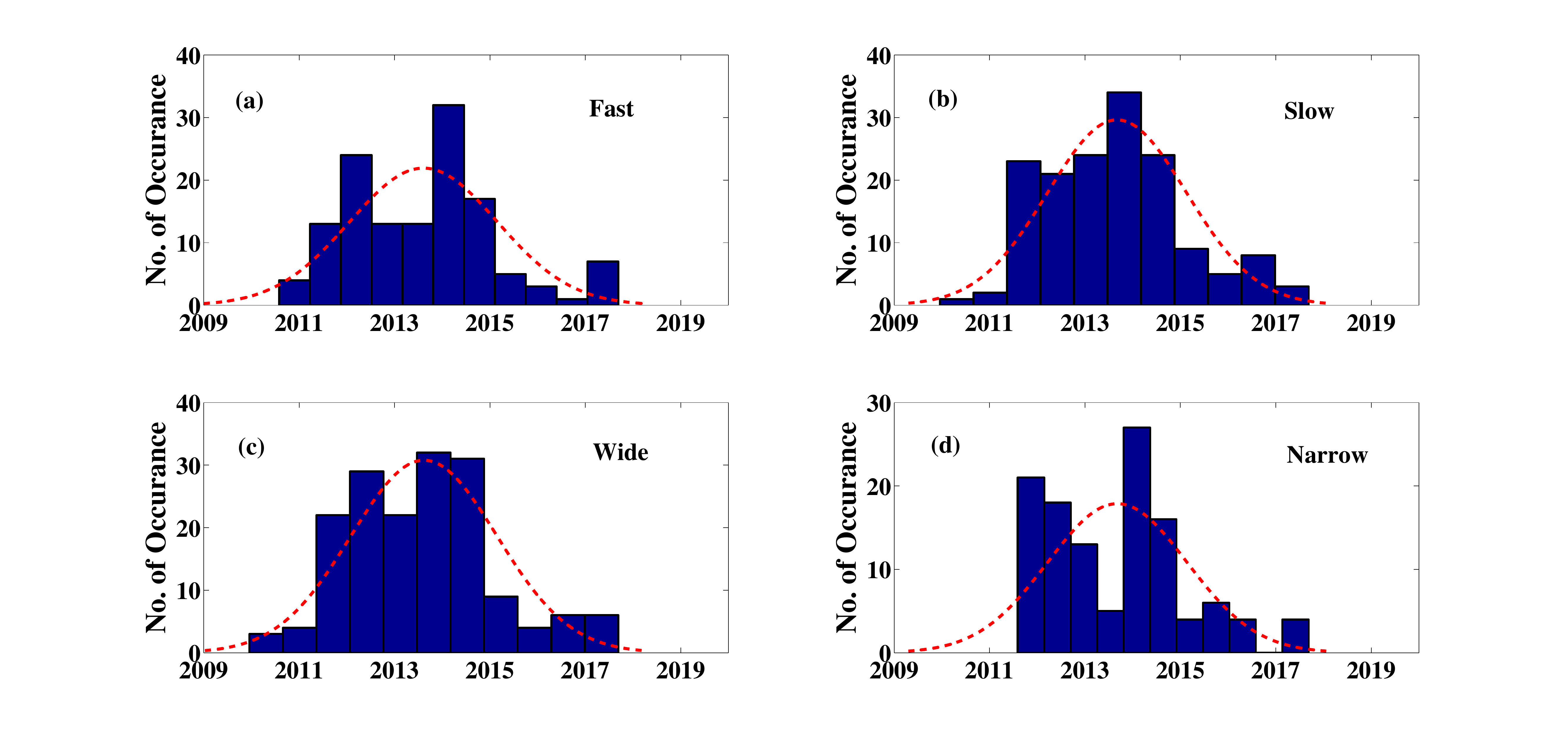}}
   \centerline{\includegraphics[width=.9\textwidth,clip=]{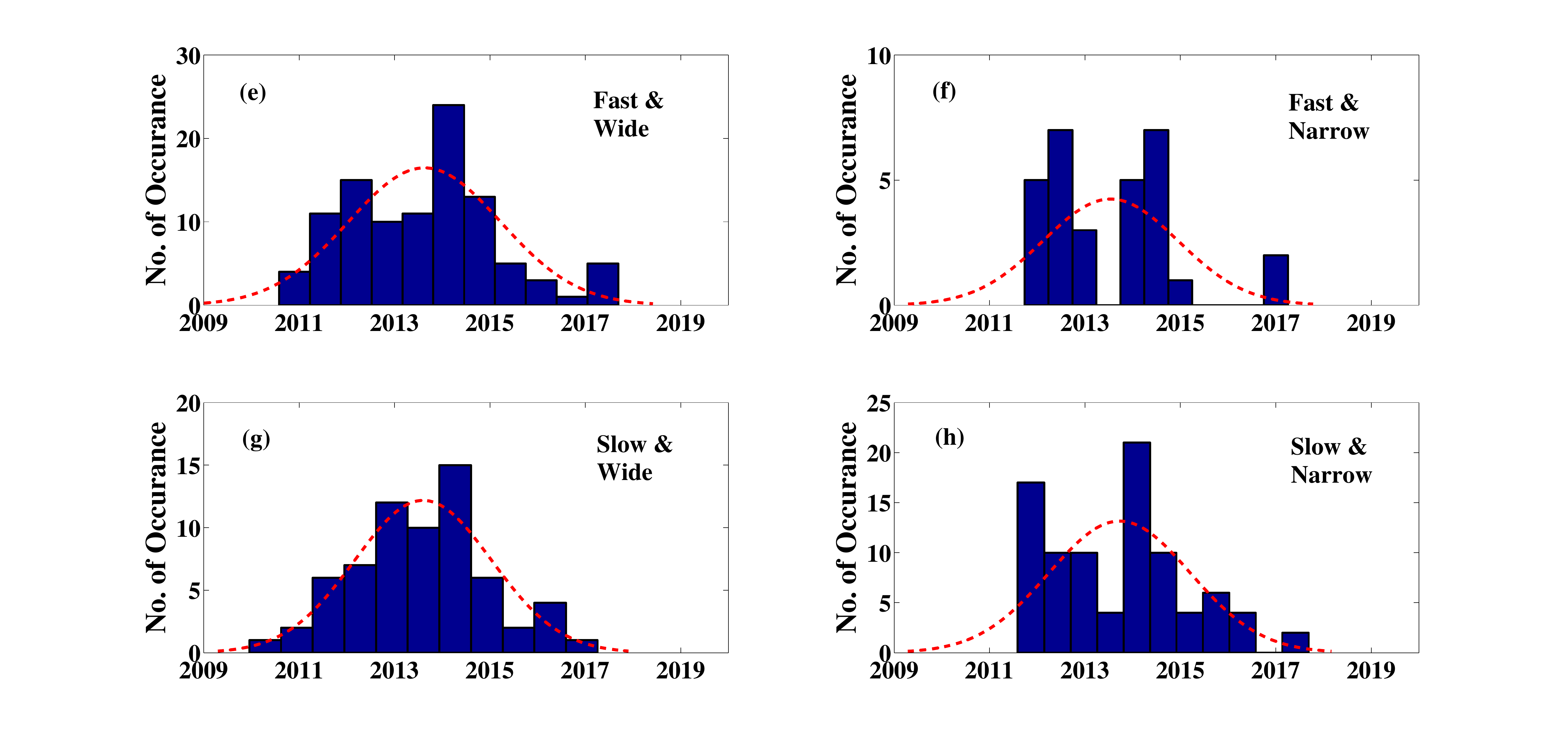}}
              \caption{  The variation of angular width and linear speed of CMEs which were accompanied with type IVs bursts in solar cycle 24. Histogram for
              (a) Fast CMEs;
              (b) Slow CMEs;
              (c) Wide CMEs;
              (d) Narrow CMEs;
              (e) Fast and Wide CMEs;
              (f) Fast and Narrow CMEs;
              (g) Slow and Wide CMEs;
              (h) Slow and Narrow CMEs.
                      }
   \label{fig:figure5b}
   \end{figure}

A distribution of stationary type IV (type IVs) bursts associated with CMEs is shown in Figure~\ref{fig:figure1}(e). Out of 366 type IVs bursts, 286 bursts ($\sim78 \%$) were accompanied with white-light CMEs. Only $\sim1.8\%$ of the total CMEs in this solar cycle were accompanied with type IVs bursts. Table \ref{tab:table2} contains the list of type IVs bursts which were associated with `Fast', `Slow', `Wide' and `Narrow' CMEs, similar to our analysis of moving type IV bursts in the previous section. 
Type IVs bursts are almost equally associated with `Fast' and `Slow' CMEs, with $\sim 46\%$ and $\sim 54\%$, respectively, and also have quite similar associations with 'Wide' and 'Narrow' CMEs, with $\sim 59\%$ and $\sim 41\%$, respectively.
Figure \ref{fig:figure5b} shows histograms of the distribution of type IVs bursts associated with various CME types. The spread of type IV bursts in this case also follows a Gaussian distribution with peaks during January to June 2014 (during the solar maximum). 
Our analysis shows that the type IVs bursts can be associated with CMEs of any speed and width.
There are very few type IVs bursts ($\sim 10 \%$) associated with `Fast and Narrow' CMEs (see Figure \ref{fig:figure5b}f) similar to type IVm bursts. However, almost $\sim 35\%$ and $\sim 30\%$ of the type IVs bursts associated with `Fast and Wide' and `Slow and Narrow' CMEs, respectively. 
In general, stationary type IV bursts show a more uniform distribution with different types of CMEs compared to moving type IV bursts.

\subsection{Type IV bursts and their properties}
\label{sec:section4.3}

\begin{figure}    
  \centering\includegraphics[width=0.32\textwidth,clip=]{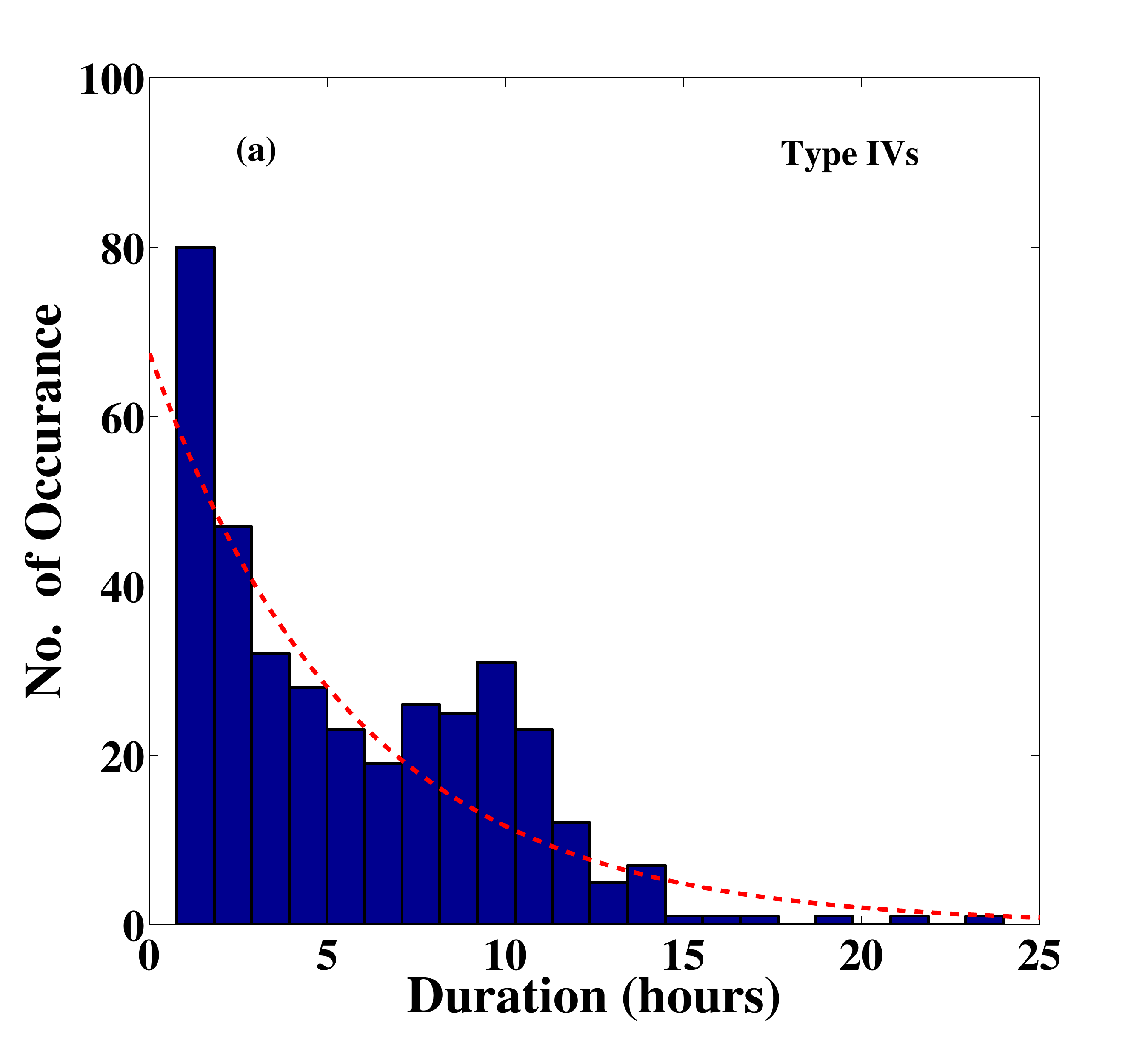}
 \centering \includegraphics[width=0.32\textwidth,clip=]{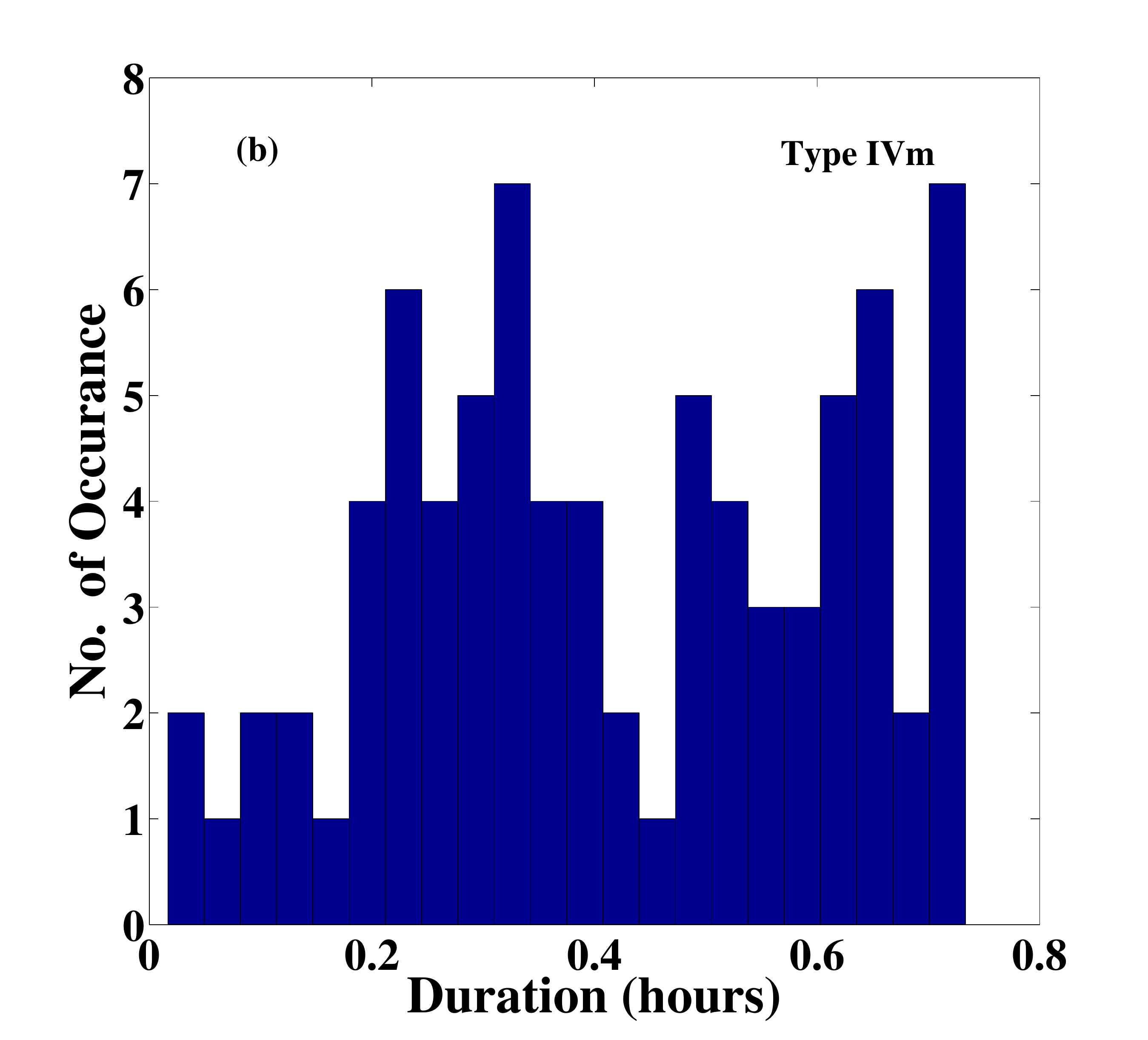}
 \centering \includegraphics[width=0.32\textwidth,clip=]{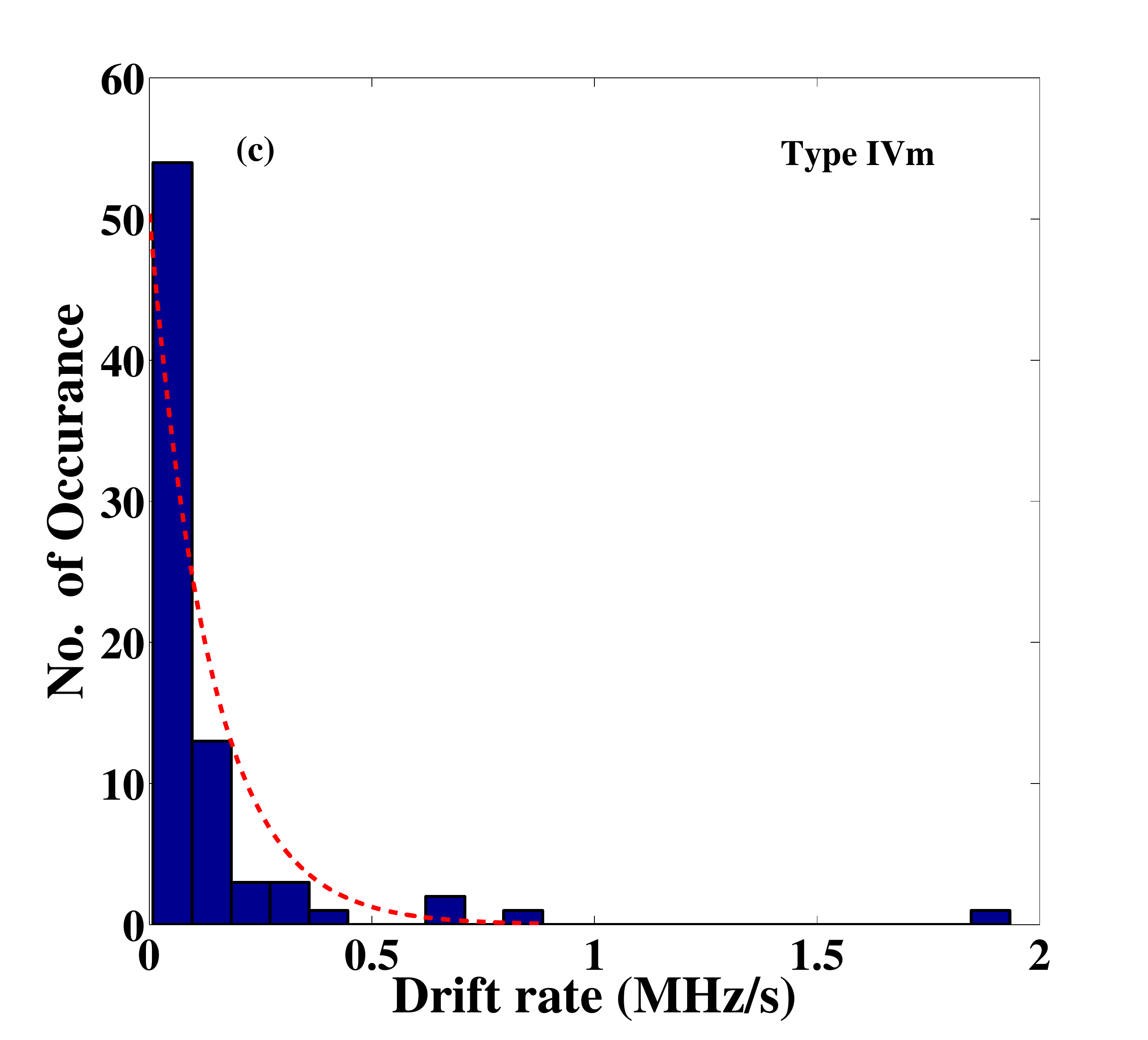}
      \caption        { Histogram showing the variation of 
              (a) the type IVs bursts duration in this cycle; 
              (b) the type IVm bursts duration in this cycle. Most of the bursts lasted for less than an hour. 
              (c) the drift rates type IVm bursts in solar cycle 24.
                      }
   \label{fig:figure6}
   \end{figure}

We analyse the typical frequency--time characteristics of type IVm and type IVs bursts using the radio spectra as shown in Figure \ref{fig:figure1a}. Figure \ref{fig:figure6} shows the variation in duration of type IVm and IVs bursts. Most of the type IVm bursts ($\sim85 \%$) had duration of a few minutes. Contrary to this, the duration of type IVs varies from a few minutes to a few hours. 
This agrees with previous observations such as the type IV bursts analyzed by \citet{pick1986observations} and \citet{melnik2018solar} which lasted up to several hours. \citet{salas2020polarisation} also studied stationary type IV bursts with durations of  $\sim 7$ hours. It is noteworthy to mention that in our study $\sim2\%$ of type IVs bursts had a duration longer than 14 hours. 

Our analysis shows that $\sim 95 \%$ of type IVm bursts had drift rates $\leq 0.5$ MHz/s, 
Similar values were found by \citet{melnik2018solar} in the range $\sim 1-2$ MHz/s. 
Figure \ref{fig:figure6}c shows the drift rate of type IVm bursts, where the red dashed line represents an exponential fit to the distribution.

\section{Discussions and Conclusions}
\label{sec:section5}

\begin{figure}    
 \centering\includegraphics[width=0.4\textwidth,clip=]{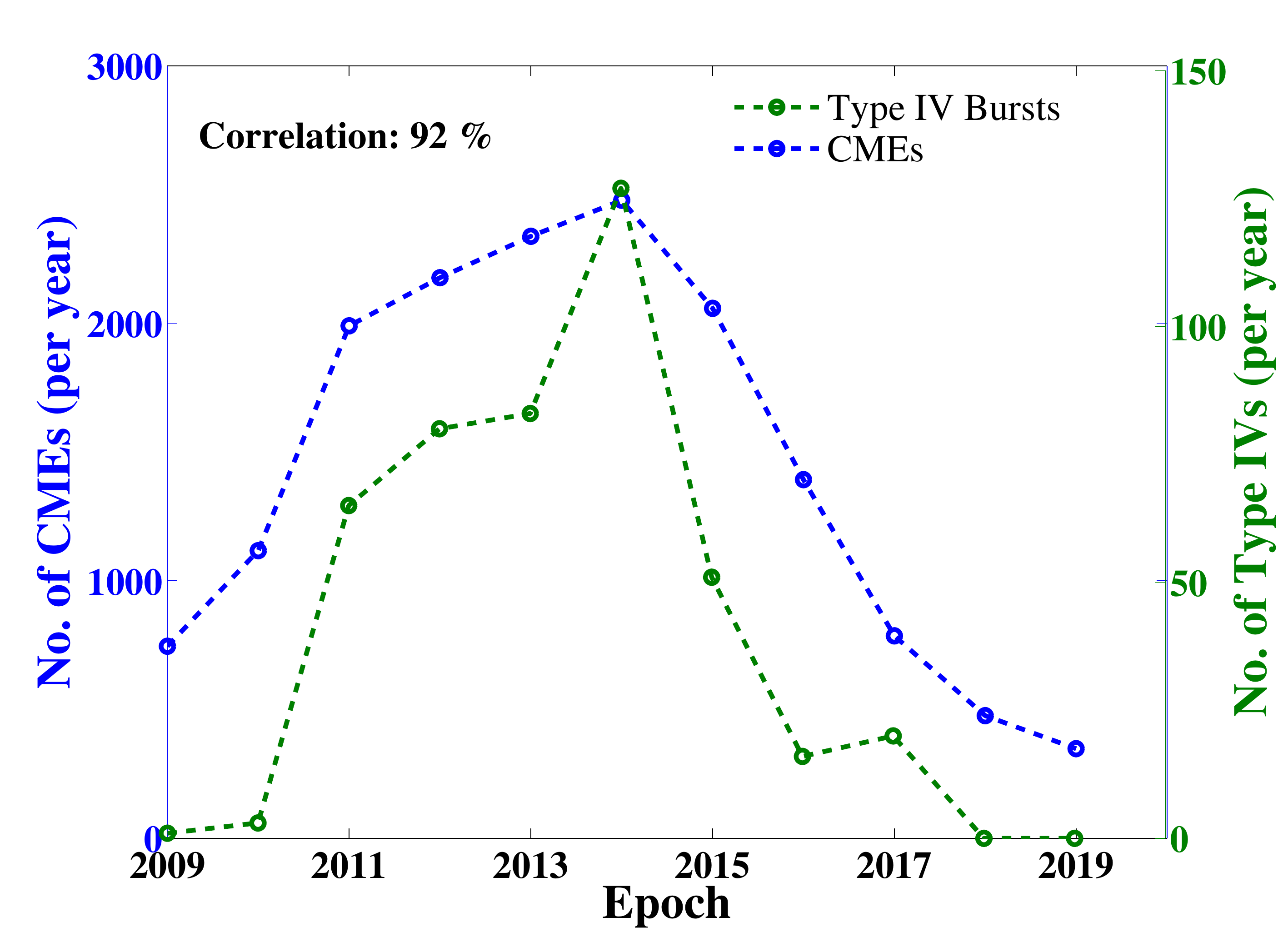}
              \caption{ The number of CMEs (per year) and type IV bursts (per year) from 2009-2019. The number of type IV bursts are very less as compared to the number of CMEs however, there is a very strong correlation ($92 \%$) between the occurrence of the the white-light feature and radio emission. 
              }
   \label{fig:figure8}
   \end{figure}

In this paper, we present the first comprehensive long-term statistical study of type IV radio bursts during solar cycle 24, where we find a significant correlation between the occurrence of type IV radio bursts and CMEs, with $\sim$81\% of the observed bursts being accompanied by CMEs. The remaining $\sim 19\%$ of type IV bursts (or 87 type IV events in total) do not show a clear CME association. Only $\sim 18 \%$ of all bursts were type IVm bursts. Almost all the type IVm bursts ($\sim91 \%$) were associated with CMEs while for type IVs bursts the association was lower, $\sim78 \%$. 
The yearly number of CMEs and type IV bursts which occurred during the solar cycle 24 is shown in Figure \ref{fig:figure8}. The correlation between the occurrence of these two events is $\sim 92 \%$, which indicates that the occurrence of type IV bursts closely follows the occurrence of CMEs. Thus, we conclude that a CME eruption may be necessary for the generation of Type IV emission as opposed to non--eruptive flares.

Further investigations are needed to determine the source of type IV bursts not associated with white-light CMEs, since they represent $\sim 19\%$ (80) of the total number of type IV bursts. From these, 92\% (80) are stationary and 8\% (7) are moving type IV bursts. One possible cause of this emission are stealth CMEs, which generally lack low coronal signatures \citep[for example,][]{Robbrecht2009,Dhuys2014A}, but are sometimes also very difficult to detect or are practically not visible in white-light signatures, even when viewed from the side \citep[][]{Kilpua2014}. It is however not clear how frequent the latter cases are.  

For those type IV bursts that are not related to CMEs, we checked if they had a flare association. To do this, we compared the start time of the type IV bursts with the start and/or peak time of flares. We found that only $\sim 16 \%$ of the total number of type IVs had a flare association without a CME association. 
In these cases, the emission can originate from electrons trapped inside post eruption flare loops that are accelerated during flaring processes \citep{Morosan2019}. Therefore, a CME eruption is not always needed to justify type IV emission and type IV emissions can also act as an important tool to diagnose the electron acceleration in the solar corona if there are no CMEs present. The above result also indicates that that there is a non-negligible number of type IVs (14 in total in our study) that lack both CME and flare association. These type IV bursts were found to occur either during the decaying phase of the flare or associated with post-eruption loops \citep{Morosan2019}.


Our analysis shows that moving type IV emission were more frequently associated with `fast' rather than `slow' CMEs and with `wide' rather than narrow CMEs. In turn, the occurrence of stationary type IV bursts did not depend on CME speed, but they were clearly more frequently associated with `wide' CMEs than narrow ones (see Table \ref{tab:table2} for details).
The combination of fast and wide CME was most likely to produce moving type IV bursts ($\sim 52 \%$ of type IVm bursts were associated with these CMEs), while fast and narrow CMEs were the least likely association (only in $\sim 10 \%$ of cases). Interestingly, a much larger fraction ($\sim 21 \%$) of moving type IV bursts were associated with CMEs that were slow and narrow, which are generally considered less significant, in particular from the space weather perspective. 





\begin{table}[]
\caption{The typical parameters type IV bursts in solar cycle 24}
\label{tab:table4}
\begin{tabular}{cccccccc}
\hline
\multirow{2}{*}{Parameter}                              & \multirow{2}{*}{}             & \multirow{2}{*}{Type IV}                 & \multirow{2}{*}{Type IVm} & \multirow{2}{*}{Type IVs} & Type IV              & Type IVm             & Type IVs             \\
                                                        &                               &                                          &                           &                           & with CME             & with CME             & with CME             \\ \hline
\multicolumn{1}{c}{\multirow{4}{*}{Duration (hh)}}      & \multicolumn{1}{c}{Mean}      & \multicolumn{1}{c}{4.7}                  & 0.4                       & 5.7                       & 4.5                  & 0.2                  & 5.5                  \\
\multicolumn{1}{c}{}                                    & \multicolumn{1}{c}{Median}    & \multicolumn{1}{c}{3.2}                  & 0.4                       & 4.7                       & 2.7                  & 0.4                  & 4.3                  \\
\multicolumn{1}{c}{}                                    & \multicolumn{1}{c}{Standard}  & \multicolumn{1}{c}{\multirow{2}{*}{4.2}} & \multirow{2}{*}{0.2}      & \multirow{2}{*}{4.1}      & \multirow{2}{*}{4.3} & \multirow{2}{*}{0.2} & \multirow{2}{*}{4.2} \\
\multicolumn{1}{c}{}                                    & \multicolumn{1}{c}{Deviation} & \multicolumn{1}{c}{}                     &                           &                           &                      &                      &                      \\ \hline
\multicolumn{1}{c}{\multirow{4}{*}{Drift Rate (MHz/s)}} & \multicolumn{1}{c}{Mean}      & \multicolumn{1}{c}{-}                    & 0.1                       & -                         & -                    & 0.1                  & -                    \\
\multicolumn{1}{c}{}                                    & \multicolumn{1}{c}{Median}    & \multicolumn{1}{c}{-}                    & 0.1                       & -                         & -                    & 0.1                  & -                    \\
\multicolumn{1}{c}{}                                    & \multicolumn{1}{c}{Standard}  & \multicolumn{1}{c}{\multirow{2}{*}{-}}   & \multirow{2}{*}{0.2}      & \multirow{2}{*}{-}        & \multirow{2}{*}{-}   & \multirow{2}{*}{0.2} & \multirow{2}{*}{-}   \\
\multicolumn{1}{c}{}                                    & \multicolumn{1}{c}{Deviation} & \multicolumn{1}{c}{}                     &                           &                           &                      &                      &                      \\ \hline
\multirow{4}{*}{Speed (km/s)}                           & Mean                          & -                                        & -                         & -                         & 684                  & 876                  & 634                  \\
                                                        & Median                        & -                                        & -                         & -                         & 494                  & 583                  & 475                  \\
                                                        & Standard                      & \multirow{2}{*}{-}                       & \multirow{2}{*}{-}        & \multirow{2}{*}{-}        & \multirow{2}{*}{599} & \multirow{2}{*}{743} & \multirow{2}{*}{546} \\
                                                        & Deviation                     &                                          &                           &                           &                      &                      &                      \\ \hline
\multirow{4}{*}{Width  (deg)}                           & Mean                          & -                                        & -                         & -                         & 135                  & 156                  & 129                  \\
                                                        & Median                        & -                                        & -                         & -                         & 82                   & 119                  & 75                   \\
                                                        & Standard                      & \multirow{2}{*}{-}                       & \multirow{2}{*}{-}        & \multirow{2}{*}{-}        & \multirow{2}{*}{122} & \multirow{2}{*}{123} & \multirow{2}{*}{121} \\ 
                                                        & Deviation                     &                                          &                           &                           &                      &                      &       
                                                        \\ \hline
\end{tabular}
\end{table}

This study also provides the typical duration of type IV bursts and drift rates of moving type IV bursts. Table \ref{tab:table4} lists the typical mean, median and standard deviation values of various parameters associated with type IV bursts.
The average duration for all the type IV bursts and type IV with CMEs were almost same as $\approx$ 4.7 hours for the former case and $\approx$ 4.5 hours for the latter case. 
The mean duration for all the type IVs bursts and type IVs with CMEs were also almost the same. 
However, the longest duration moving type IV bursts were found for cases not associated with CMEs.
The mean drift rate for all moving type IV bursts and the moving type IV bursts associated with white-light CMEs were the same ($\approx$ 0.1 MHz/s). 
Moving type IV bursts associated with CMEs typically have shorter duration but almost the same drift rate as the moving type IV bursts without CMEs. 
Note that these values have high standard deviations. 
The mean speed of CMEs associated with stationary type IV bursts was higher than that of moving type IV bursts. This is likely due to the fact that stationary type IV bursts are almost equally associated with `Fast' and `Slow' CMEs.
Moving type IV bursts were associated with wider CMEs as compared to stationary type IV bursts.

The high association of type IV bursts with CMEs indicates that type IV, when present, can be used for studying the electron acceleration locations as well as CME kinematics early during the eruption process.  
We used the duration of the type IV bursts in the spectra as a classifier for moving/stationary bursts in the absence of imaging observations. We also checked the drift rates of the moving bursts and it was found that the drift rates of most of the moving bursts classified this way were $\geq 0.03$ MHz. To confirm whether a radio burst is a moving or not, usually radio imaging observations are needed.
We note that due to the absence of radio imaging, a few type IV bursts were put into wrong groups during the event classification. For example, we used the duration of the radio burst as first stage classification based on the work by  \cite{Robinson1978} and \cite{Gergely1986}. There have been two such reported case studies in the recent past, which showed that moving type IV bursts can have duration $>1$ hours sometimes \citep{Ramesh2013, Vasanth2019}. In the duration/drift rate based automated classification in this study, there bursts were classified as stationary burst because of their spectral properties.
Our classification relies upon the spectral data however, this is the best approach study that can be done given the instrumental capabilities during the solar Cycle 24 and the lack of radio images available for the majority of these bursts.
Our work outline the necessity of radio imaging to provide the possibility of identifying the source region of type IV emission and electron acceleration in relation to the accompanying CME or flare.
This study will greatly benefit from future investigations using instruments capable of continuous imaging  and spectroscopic observations in the radio domain. 


\acknowledgments

The CME catalog is generated and maintained at the CDAW Data Center by NASA and The Catholic University of America in cooperation with the Naval Research Laboratory. SOHO is a project of international cooperation between ESA and NASA. We acknowledge Space Weather Prediction Center, Boulder for event lists. The sunspot data was taken from the World Data Center SILSO, Royal Observatory of Belgium, Brussels.
D.E.M. acknowledges the Academy of Finland Project 333859 and the Finnish Centre of Excellence in Research of Sustainable Space (Academy of Finland grant number 312390). A.K. and E.K.J.K. acknowledge the ERC under the European Union's Horizon 2020 Research and Innovation Programme Project SolMAG 724391. E.K.J.K. also acknowledges the Academy of Finland Project 310445. 

\bibliography{reference}{}

\begin{thebibliography}{}
\expandafter\ifx\csname natexlab\endcsname\relax\def\natexlab#1{#1}\fi
\providecommand{\url}[1]{\href{#1}{#1}}
\providecommand{\dodoi}[1]{doi:~\href{http://doi.org/#1}{\nolinkurl{#1}}}
\providecommand{\doeprint}[1]{\href{http://ascl.net/#1}{\nolinkurl{http://ascl.net/#1}}}
\providecommand{\doarXiv}[1]{\href{https://arxiv.org/abs/#1}{\nolinkurl{https://arxiv.org/abs/#1}}}

\bibitem[{{Bain} {et~al.}(2014){Bain}, {Krucker}, {Saint-Hilaire}, \&
  {Raftery}}]{Bain2014}
{Bain}, H.~M., {Krucker}, S., {Saint-Hilaire}, P., \& {Raftery}, C.~L. 2014,
  \apj, 782, 43, \dodoi{10.1088/0004-637X/782/1/43}

\bibitem[{{Bastian} {et~al.}(1998){Bastian}, {Benz}, \& {Gary}}]{Bastian1998}
{Bastian}, T.~S., {Benz}, A.~O., \& {Gary}, D.~E. 1998, \araa, 36, 131,
  \dodoi{10.1146/annurev.astro.36.1.131}

\bibitem[{{Bastian} {et~al.}(2001){Bastian}, {Pick}, {Kerdraon}, {Maia}, \&
  {Vourlidas}}]{Bastian2001}
{Bastian}, T.~S., {Pick}, M., {Kerdraon}, A., {Maia}, D., \& {Vourlidas}, A.
  2001, \apjl, 558, L65, \dodoi{10.1086/323421}

\bibitem[{{Benz} \& {Tarnstrom}(1976)}]{Benz1976}
{Benz}, A.~O., \& {Tarnstrom}, G.~L. 1976, \apj, 204, 597,
  \dodoi{10.1086/154208}

\bibitem[{{Boischot}(1957)}]{Boischot1957}
{Boischot}, A. 1957, Academie des Sciences Paris Comptes Rendus, 244, 1326

\bibitem[{{Boischot} \& {Clavelier}(1968)}]{Boischot1968}
{Boischot}, A., \& {Clavelier}, B. 1968, in IAU Symposium, Vol.~35, Structure
  and Development of Solar Active Regions, ed. K.~O. {Kiepenheuer}, 565

\bibitem[{{Brueckner} {et~al.}(1995){Brueckner}, {Howard}, {Koomen},
  {Korendyke}, {Michels}, {Moses}, {Socker}, {Dere}, {Lamy}, {Llebaria},
  {Bout}, {Schwenn}, {Simnett}, {Bedford}, \& {Eyles}}]{Brueckner1995}
{Brueckner}, G.~E., {Howard}, R.~A., {Koomen}, M.~J., {et~al.} 1995, \solphys,
  162, 357, \dodoi{10.1007/BF00733434}

\bibitem[{{Carley} {et~al.}(2017){Carley}, {Vilmer}, {Sim{\~o}es}, \& {{\'O}
  Fearraigh}}]{Carley2017}
{Carley}, E.~P., {Vilmer}, N., {Sim{\~o}es}, P.~J.~A., \& {{\'O} Fearraigh}, B.
  2017, \aap, 608, A137, \dodoi{10.1051/0004-6361/201731368}

\bibitem[{{Cla{\ss}en} \& {Aurass}(2002)}]{Clasen2002}
{Cla{\ss}en}, H.~T., \& {Aurass}, H. 2002, \aap, 384, 1098,
  \dodoi{10.1051/0004-6361:20020092}

\bibitem[{{D'Huys} {et~al.}(2014){D'Huys}, {Seaton}, {Poedts}, \&
  {Berghmans}}]{Dhuys2014A}
{D'Huys}, E., {Seaton}, D.~B., {Poedts}, S., \& {Berghmans}, D. 2014, \apj,
  795, 49, \dodoi{10.1088/0004-637X/795/1/49}

\bibitem[{{Dulk}(1973)}]{Dulk1973}
{Dulk}, G.~A. 1973, \solphys, 32, 491, \dodoi{10.1007/BF00154962}

\bibitem[{{Gary} {et~al.}(1985){Gary}, {Dulk}, {House}, {Illing}, \&
  {Wagner}}]{Gary1985}
{Gary}, D.~E., {Dulk}, G.~A., {House}, L.~L., {Illing}, R., \& {Wagner}, W.~J.
  1985, \aap, 152, 42

\bibitem[{{Gergely}(1986)}]{Gergely1986}
{Gergely}, T.~E. 1986, \solphys, 104, 175, \dodoi{10.1007/BF00159959}

\bibitem[{{Gopalswamy} \& {Kundu}(1987)}]{Gopal1987}
{Gopalswamy}, N., \& {Kundu}, M.~R. 1987, \solphys, 114, 347,
  \dodoi{10.1007/BF00167350}

\bibitem[{Gopalswamy {et~al.}(2004)Gopalswamy, Yashiro, Krucker, Stenborg, \&
  Howard}]{gopalswamy2004intensity}
Gopalswamy, N., Yashiro, S., Krucker, S., Stenborg, G., \& Howard, R.~A. 2004,
  Journal of Geophysical Research: Space Physics, 109

\bibitem[{{Gopalswamy} {et~al.}(2009){Gopalswamy}, {Yashiro}, {Michalek},
  {Stenborg}, {Vourlidas}, {Freeland}, \& {Howard}}]{Gopalswamy2009cdaw}
{Gopalswamy}, N., {Yashiro}, S., {Michalek}, G., {et~al.} 2009, Earth Moon and
  Planets, 104, 295, \dodoi{10.1007/s11038-008-9282-7}

\bibitem[{{Hariharan} {et~al.}(2016){Hariharan}, {Ramesh}, {Kathiravan}, \&
  {Wang}}]{Hariharan2016a}
{Hariharan}, K., {Ramesh}, R., {Kathiravan}, C., \& {Wang}, T.~J. 2016,
  \solphys, 291, 1405, \dodoi{10.1007/s11207-016-0918-x}

\bibitem[{{Howard} {et~al.}(2008){Howard}, {Moses}, {Vourlidas}, {Newmark},
  {Socker}, {Plunkett}, {Korendyke}, {Cook}, {Hurley}, {Davila}, {Thompson},
  {St Cyr}, {Mentzell}, {Mehalick}, {Lemen}, {Wuelser}, {Duncan}, {Tarbell},
  {Wolfson}, {Moore}, {Harrison}, {Waltham}, {Lang}, {Davis}, {Eyles},
  {Mapson-Menard}, {Simnett}, {Halain}, {Defise}, {Mazy}, {Rochus}, {Mercier},
  {Ravet}, {Delmotte}, {Auchere}, {Delaboudiniere}, {Bothmer}, {Deutsch},
  {Wang}, {Rich}, {Cooper}, {Stephens}, {Maahs}, {Baugh}, {McMullin}, \&
  {Carter}}]{Howard2008}
{Howard}, R.~A., {Moses}, J.~D., {Vourlidas}, A., {et~al.} 2008, \ssr, 136, 67,
  \dodoi{10.1007/s11214-008-9341-4}

\bibitem[{{Kahler} {et~al.}(2019){Kahler}, {Ling}, \&
  {Gopalswamy}}]{Kahler2019}
{Kahler}, S.~W., {Ling}, A.~G., \& {Gopalswamy}, N. 2019, \solphys, 294, 134,
  \dodoi{10.1007/s11207-019-1518-3}

\bibitem[{{Kilpua} {et~al.}(2017){Kilpua}, {Balogh}, {von Steiger}, \&
  {Liu}}]{Kilpua2017}
{Kilpua}, E.~K.~J., {Balogh}, A., {von Steiger}, R., \& {Liu}, Y.~D. 2017,
  \ssr, 212, 1271, \dodoi{10.1007/s11214-017-0411-3}

\bibitem[{{Kilpua} {et~al.}(2019){Kilpua}, {Lugaz}, {Mays}, \&
  {Temmer}}]{Kilpua2019}
{Kilpua}, E.~K.~J., {Lugaz}, N., {Mays}, M.~L., \& {Temmer}, M. 2019, Space
  Weather, 17, 498, \dodoi{10.1029/2018SW001944}

\bibitem[{{Kilpua} {et~al.}(2014){Kilpua}, {Mierla}, {Zhukov}, {Rodriguez},
  {Vourlidas}, \& {Wood}}]{Kilpua2014}
{Kilpua}, E.~K.~J., {Mierla}, M., {Zhukov}, A.~N., {et~al.} 2014, \solphys,
  289, 3773, \dodoi{10.1007/s11207-014-0552-4}

\bibitem[{{Koval} {et~al.}(2016){Koval}, {Stanislavsky}, {Chen}, {Feng},
  {Konovalenko}, \& {Volvach}}]{Koval2016}
{Koval}, A., {Stanislavsky}, A., {Chen}, Y., {et~al.} 2016, \apj, 826, 125,
  \dodoi{10.3847/0004-637X/826/2/125}

\bibitem[{{Kumari} {et~al.}(2017{\natexlab{a}}){Kumari}, {Ramesh},
  {Kathiravan}, \& {Gopalswamy}}]{Anshu2017a}
{Kumari}, A., {Ramesh}, R., {Kathiravan}, C., \& {Gopalswamy}, N.
  2017{\natexlab{a}}, \apj, 843, 10, \dodoi{10.3847/1538-4357/aa72e7}

\bibitem[{{Kumari} {et~al.}(2017{\natexlab{b}}){Kumari}, {Ramesh},
  {Kathiravan}, \& {Wang}}]{kumari2017b}
{Kumari}, A., {Ramesh}, R., {Kathiravan}, C., \& {Wang}, T.~J.
  2017{\natexlab{b}}, Solar Physics, 292, 161,
  \dodoi{10.1007/s11207-017-1180-6}

\bibitem[{{Kumari} {et~al.}(2019){Kumari}, {Ramesh}, {Kathiravan}, {Wang}, \&
  {Gopalswamy}}]{kumari2019direct}
{Kumari}, A., {Ramesh}, R., {Kathiravan}, C., {Wang}, T.~J., \& {Gopalswamy},
  N. 2019, The Astrophysical Journal, 881, 24, \dodoi{10.3847/1538-4357/ab2adf}

\bibitem[{Lara {et~al.}(2003)Lara, Gopalswamy, Nunes, Munoz, \&
  Yashiro}]{lara2003statistical}
Lara, A., Gopalswamy, N., Nunes, S., Munoz, G., \& Yashiro, S. 2003,
  Geophysical Research Letters, 30

\bibitem[{{Liu} {et~al.}(2018){Liu}, {Chen}, {Cho}, {Feng}, {Vasanth}, {Koval},
  {Du}, {Wu}, \& {Li}}]{Liu2018}
{Liu}, H., {Chen}, Y., {Cho}, K., {et~al.} 2018, \solphys, 293, 58,
  \dodoi{10.1007/s11207-018-1280-y}

\bibitem[{{Maia} {et~al.}(2007){Maia}, {Gama}, {Mercier}, {Pick}, {Kerdraon},
  \& {Karlick{\'y}}}]{Maia2007}
{Maia}, D.~J.~F., {Gama}, R., {Mercier}, C., {et~al.} 2007, \apj, 660, 874,
  \dodoi{10.1086/508011}

\bibitem[{{Melnik} {et~al.}(2018){Melnik}, {Rucker}, {Konovalenko},
  {Dorovskyy}, {Abranin}, {Brazhenko}, {Thide}, \&
  {Stanislavskyy}}]{melnik2018solar}
{Melnik}, V.~N., {Rucker}, H.~O., {Konovalenko}, A.~A., {et~al.} 2018, arXiv
  e-prints, arXiv:1802.06249.
\newblock \doarXiv{1802.06249}

\bibitem[{Melrose(1980)}]{melrose1980emission}
Melrose, D. 1980, Space Science Reviews, 26, 3

\bibitem[{{Morosan} {et~al.}(2019){Morosan}, {Kilpua}, {Carley}, \&
  {Monstein}}]{Morosan2019}
{Morosan}, D.~E., {Kilpua}, E.~K.~J., {Carley}, E.~P., \& {Monstein}, C. 2019,
  \aap, 623, A63, \dodoi{10.1051/0004-6361/201834510}

\bibitem[{{Morosan, D. E.} {et~al.}(2020){Morosan, D. E.}, {Palmerio, E.},
  {Pomoell, J.}, {Vainio, R.}, {Palmroth, M.}, \& {Kilpua, E. K.
  J.}}]{Morosan2020}
{Morosan, D. E.}, {Palmerio, E.}, {Pomoell, J.}, {et~al.} 2020, A\&A, 635, A62,
  \dodoi{10.1051/0004-6361/201937133}

\bibitem[{{Pick}(1986)}]{pick1986observations}
{Pick}, M. 1986, \solphys, 104, 19, \dodoi{10.1007/BF00159942}

\bibitem[{{Ramesh} {et~al.}(2004){Ramesh}, {Kathiravan}, \& {Satya
  Narayanan}}]{Ramesh2004}
{Ramesh}, R., {Kathiravan}, C., \& {Satya Narayanan}, A. 2004, Asian Journal of
  Physics, 13, 277

\bibitem[{{Ramesh} {et~al.}(2013){Ramesh}, {Kishore}, {Mulay}, {Barve},
  {Kathiravan}, \& {Wang}}]{Ramesh2013}
{Ramesh}, R., {Kishore}, P., {Mulay}, S.~M., {et~al.} 2013, \apj, 778, 30,
  \dodoi{10.1088/0004-637X/778/1/30}

\bibitem[{{Robbrecht} {et~al.}(2009){Robbrecht}, {Patsourakos}, \&
  {Vourlidas}}]{Robbrecht2009}
{Robbrecht}, E., {Patsourakos}, S., \& {Vourlidas}, A. 2009, \apj, 701, 283,
  \dodoi{10.1088/0004-637X/701/1/283}

\bibitem[{{Robinson}(1978)}]{Robinson1978}
{Robinson}, R.~D. 1978, \solphys, 60, 383, \dodoi{10.1007/BF00156538}

\bibitem[{{Salas-Matamoros} \& {Klein}(2020)}]{salas2020polarisation}
{Salas-Matamoros}, C., \& {Klein}, K.-L. 2020, \aap, 639, A102,
  \dodoi{10.1051/0004-6361/202037989}

\bibitem[{Schmahl(1973)}]{schmahl1973}
Schmahl, E. 1973, Australian Journal of Physics Astrophysical Supplement, 29, 1

\bibitem[{{Tun} \& {Vourlidas}(2013)}]{Tun2013}
{Tun}, S.~D., \& {Vourlidas}, A. 2013, \apj, 766, 130,
  \dodoi{10.1088/0004-637X/766/2/130}

\bibitem[{Vasanth {et~al.}(2016)Vasanth, Chen, Feng, Ma, Du, Song, Kong, \&
  Wang}]{Vasanth_2016}
Vasanth, V., Chen, Y., Feng, S., {et~al.} 2016, The Astrophysical Journal, 830,
  L2, \dodoi{10.3847/2041-8205/830/1/l2}

\bibitem[{Vasanth {et~al.}(2019)Vasanth, Chen, Lv, Ning, Li, Feng, Wu, \&
  Du}]{Vasanth2019}
Vasanth, V., Chen, Y., Lv, M., {et~al.} 2019, The Astrophysical Journal, 870,
  30, \dodoi{10.3847/1538-4357/aaeffd}

\bibitem[{{Vourlidas} {et~al.}(2017){Vourlidas}, {Balmaceda}, {Stenborg}, \&
  {Dal Lago}}]{Vourlidas2017}
{Vourlidas}, A., {Balmaceda}, L.~A., {Stenborg}, G., \& {Dal Lago}, A. 2017,
  \apj, 838, 141, \dodoi{10.3847/1538-4357/aa67f0}

\bibitem[{Webb \& Howard(2012)}]{webb2012coronal}
Webb, D.~F., \& Howard, T.~A. 2012, Living Reviews in Solar Physics, 9, 3

\bibitem[{{Weiss}(1963)}]{Weiss1963}
{Weiss}, A.~A. 1963, Australian Journal of Physics, 16, 526,
  \dodoi{10.1071/PH630526}

\bibitem[{White(2007)}]{white2007solar}
White, S.~M. 2007, Asian Journal of Physics, 16, 189

\bibitem[{{Yashiro} {et~al.}(2004){Yashiro}, {Gopalswamy}, {Michalek}, {St.
  Cyr}, {Plunkett}, {Rich}, \& {Howard}}]{Yashiro2004}
{Yashiro}, S., {Gopalswamy}, N., {Michalek}, G., {et~al.} 2004, Journal of
  Geophysical Research (Space Physics), 109, A07105,
  \dodoi{10.1029/2003JA010282}

\bibitem[{{Yashiro} {et~al.}(2008){Yashiro}, {Michalek}, \&
  {Gopalswamy}}]{Yashiro2008}
{Yashiro}, S., {Michalek}, G., \& {Gopalswamy}, N. 2008, Annales Geophysicae,
  26, 3103, \dodoi{10.5194/angeo-26-3103-2008}

\end{thebibliography}
\bibliographystyle{aasjournal}

\end{document}